\documentclass[journal]{IEEEtran}
\usepackage{cite}
\usepackage[utf8]{inputenc}
\usepackage{graphicx}
\usepackage{amssymb}
\usepackage[usenames, dvipsnames]{xcolor}
\usepackage{footmisc}

\definecolor{twostepkcm}{RGB}{130, 130, 130}
\definecolor{twostepkcmfill}{RGB}{200, 200, 200}
\definecolor{twostepmcm}{RGB}{125, 125, 255}
\definecolor{twostepmcmfill}{RGB}{200, 200, 255}
\definecolor{twostepgeneric}{RGB}{255, 100, 100}
\definecolor{twostepgenericfill}{RGB}{255, 160, 160}
\definecolor{twostepgenericnodsp}{RGB}{255, 127, 0}
\definecolor{twostepgenericnodspfill}{RGB}{255, 227, 100}
\definecolor{ours}{RGB}{0, 150, 0}
\definecolor{oursfill}{RGB}{0, 230, 0}
\definecolor{errorplotlines}{RGB}{255, 100, 100}
\definecolor{errorplotcenterdots}{RGB}{0, 0, 0}
\definecolor{errorplotdots}{RGB}{255, 100, 100}

\usepackage{booktabs}
\usepackage{multirow}

\usepackage{pgfplots}
\pgfplotsset{compat=newest}
\usepackage{amsmath}
\usepackage{tikz}
\usepackage{url}
\usepackage{xspace}
\usepackage{tabularx}
\usepackage{hyperref}
\usepackage{booktabs}
\usepackage{enumitem}
\usepackage{multicol}
\usepackage{array}
\usepackage{lscape}
\newcolumntype{L}[1]{>{\raggedright\let\newline\\\arraybackslash\hspace{0pt}}m{#1}}
\newcolumntype{C}[1]{>{\centering\let\newline\\\arraybackslash\hspace{0pt}}m{#1}}
\newcolumntype{R}[1]{>{\raggedleft\let\newline\\\arraybackslash\hspace{0pt}}m{#1}}

\usetikzlibrary{dsp, chains, decorations.markings, positioning, shapes.geometric, shadows, fit,
backgrounds}

\pgfplotsset{
	legend entry/.initial=,
	every axis plot post/.code={%
		\pgfkeysgetvalue{/pgfplots/legend entry}\tempValue
		\ifx\tempValue\empty
		\pgfkeysalso{/pgfplots/forget plot}%
		\else
		\expandafter\addlegendentry\expandafter{\tempValue}%
		\fi
	},
}

\def\MarkLt{8pt}
\def\MarkSep{3pt}

\tikzset{
	triangleleftright/.style = {draw, regular polygon, regular polygon sides=3, shape border rotate=270},
	trianglerightleft/.style = {draw, regular polygon, regular polygon sides=3, shape border rotate=90},
	triangledownup/.style = {draw, regular polygon, regular polygon sides=3, shape border rotate=0},
	triangleupdown/.style = {draw, regular polygon, regular polygon sides=3, shape border rotate=180}
}

\tikzset{
	TwoMarks/.style={
		postaction={decorate,
			decoration={
				markings,
				mark=at position #1 with
				{
					\begin{scope}[xslant=0.2]
						\draw[line width=\MarkSep,white,-] (0pt,-\MarkLt) -- (0pt,\MarkLt);
						\draw[-] (-0.5*\MarkSep,-\MarkLt) -- (-0.5*\MarkSep,\MarkLt);
						\draw[-] (0.5*\MarkSep,-\MarkLt) -- (0.5*\MarkSep,\MarkLt);
					\end{scope}
				}
			}
		}
	},
	TwoMarks/.default={0.5},
	TwoMarksArrow/.style={
		postaction={decorate,
			decoration={
				markings,
				mark=at position #1 with
				{
					\begin{scope}[xslant=0.2]
						\draw[line width=\MarkSep,white,-] (0pt,-\MarkLt) -- (0pt,\MarkLt);
						\draw[-] (-0.5*\MarkSep,-\MarkLt) -- (-0.5*\MarkSep,\MarkLt);
						\draw[-] (0.5*\MarkSep,-\MarkLt) -- (0.5*\MarkSep,\MarkLt);
					\end{scope}
				}
			}
		}
	},
	TwoMarksArrow/.default={0.5},
	Delay/.style={
		postaction={decorate,
			decoration={
				markings,
				mark=at position #1 with
				{
					\begin{scope}
						\draw[line width=\MarkSep,white,-] (0pt,-\MarkLt) -- (0pt,\MarkLt);
						\draw[fill=black] (-0.5*\MarkSep, -\MarkLt) rectangle (0.5*\MarkSep, \MarkLt);
					\end{scope}
				}
			}
		}
	},
	Delay/.default={0.5},
	Truncature/.style={
		postaction={decorate,
			decoration={
				markings,
				mark=at position #1 with
				{
					\begin{scope}[xslant=0.3]
						\draw[-] (0,-\MarkLt) -- (0,\MarkLt);
					\end{scope}
				}
			}
		}
	},
	Truncature/.default={0.5},
}

\usepackage{subcaption}

\newcommand{\openint}[2]{\left]#1; #2\right[}

\newcommand{\closeint}[2]{\left[#1; #2\right]}

\newcommand{\sg}{\text{\normalfont sg}}

\newcommand{\outm}{\text{\normalfont out}}
\newcommand{\extm}{\text{\normalfont ext}}
\newcommand{\inm}{\text{\normalfont in}}

\newcommand{\eg}{\emph{e.\,g.}\xspace}
\newcommand{\ie}{\emph{i.\,e.}\xspace}
\newcommand{\wrt}{{w.\,r.\,t.}\xspace}
\newcommand{\st}{{s.\,t.}\xspace}
\newcommand{\filtername}[1]{\texttt{#1}} 
\newcommand{\lp}[2]{\filtername{lp#1}$_{#2}$}
\newcommand{\lpfig}[2]{lp#1$_{#2}$}

\DeclareMathOperator{\sign}{sign}

\newcommand{\TODO}[1]{\textcolor{red}{TODO: #1}}

\begin{document}

\title{Hardware-aware Design of Multiplierless Second-Order IIR Filters with Minimum Adders}
%
%
%
\author{R\'{e}mi Garcia, Anastasia Volkova, Martin Kumm, Alexandre Goldsztejn, Jonas K\"{u}hle%
\thanks{R. Garcia and A. Volkova were with Universit\'{e} de Nantes, CNRS, LS2N F-44000 Nantes,
France. Email: firstname.lastname@univ-nantes.fr}
\thanks{A. Goldsztejn was with CNRS, Universit\'{e} de Nantes, LS2N F-44000 Nantes, France. Email: alexandre.goldsztejn@ls2n.fr}
\thanks{M. Kumm and J. K\"{u}hle were with Faculty of Applied Computer Science, Fulda University of
Applied Sciences,
Germany. Email: firstname.lastname@informatik.hs-fulda.de}
}

\maketitle

\begin{abstract}
	In this work we optimally solve the problem of multiplierless design of second-order Infinite Impulse
	Response filters with minimum number of adders.
	Given a frequency specification, we design a stable direct form filter with hardware-aware fixed-point coefficients that yielding minimal number of
	adders when replacing all the multiplications by bit shifts and additions.
	The coefficient design, quantization and implementation, typically conducted independently, are now gathered into
	one global
	optimization problem, modeled through integer linear programming and efficiently solved using
	generic solvers.
	We guarantee the frequency-domain specifications and stability, which together with optimal number of adders will significantly simplify design-space exploration for filter designers.

	The optimal filters are implemented within the FloPoCo IP core generator and synthesized for Field
	Programmable Gate Arrays.
	With respect to state-of-the-art three-step filter design methods, our one-step design approach
	achieves, on average,
	42\% reduction in number of lookup tables and 21\% improvement in~delay.
\end{abstract}


%
\IEEEpeerreviewmaketitle

\section{Introduction}

\IEEEPARstart{D}{igital} filters are essential components of modern technology, from medical equipment
to scientific instruments.
Filter design is a core topic in digital signal processing and control, and efficient filter implementation in
software and hardware
has received a significant research interest for half a century.
Infinite Impulse Response (IIR) filters are a class of widely-used recursive linear time-invariant filters.
IIR filters can be relatively easily designed in software, but hardware implementation is essential for
embedded systems, where
performance/power constraints are critical. Some application domains, such as 5G/6G backbones and
autonomous vehicles, rely
on reconfigurable hardware using Field Programmable Gate Arrays (FPGA).

Classic Fixed-Point (FxP) filter design and implementation flow follows three separate steps:
\begin{enumerate}
	\item \textbf{Filter Design (FD)} consists in finding real (in practice, double precision) filter coefficients,
	adhering to the given frequency
	specification. IIR filters are defined by coefficients of a rational transfer function, for which a stability
	criteria must be also
	satisfied. In general, a large amount of different filter coefficient sets can realize a given frequency
	specification;
	\item \textbf{Quantization (Q)} converts the coefficients to a FxP format such that they still
	respect the given frequency
	response and they still lead to a stable filter;
	\item \textbf{Implementation (I)} consists in generating, using quantized coefficients, a valid hardware
	description. This step exposes a
	high number of parameters, \eg{}, the type of multipliers used.
\end{enumerate}

The combination of FD and Q steps has been studied extensively since 1960s
\cite{SmekalVich_OptimizedmodelsIIR_1999,
VanuytselBoetsVanBiesenTemmerman_Efficienthybridoptimization_2002,
WangLiLi_Fixedpointdigital_2010}. For certain structures
of IIR filters it can even be
considered solved \cite{GeversLi93}, but with respect to a signal to quantization noise ratio, which is a
probabilistic measure, not
guaranteeing numerical safety. The general approach for quantization of a transfer function is
nevertheless quite straightforward,
passing through iterative increase of coefficient word length. 

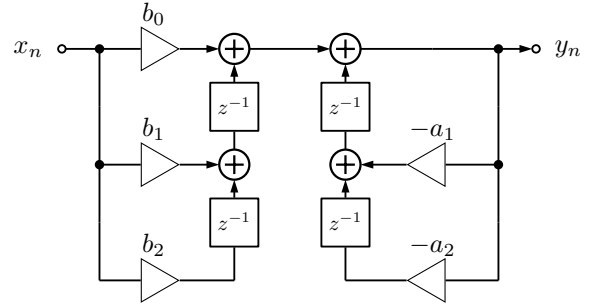
\begin{figure}
	\centering
	\begin{tikzpicture}
	\matrix (secondorderiir) [row sep=-3mm, column sep=4mm] {
		\node[dspnodeopen, dsp/label=left] (m11) {$x_n$}; \pgfmatrixnextcell
		\node[dspnodefull] (m12) {}; \pgfmatrixnextcell
		\node[triangleleftright, label={above:$b_0$}, label={below:$\phantom{b_0}$}] (m13) {}; \pgfmatrixnextcell
		\node[dspadder, label={right:$\phantom{.}$}] (m14) {}; \pgfmatrixnextcell
		\node[dspadder, label={left:$\phantom{.}$}] (m15) {}; \pgfmatrixnextcell
		\node[coordinate] (m16) {}; \pgfmatrixnextcell
		\node[dspnodefull] (m17) {}; \pgfmatrixnextcell
		\node[dspnodeopen, dsp/label=right] (m18) {$y_n$}; \\
		\node[coordinate] (m21) {}; \pgfmatrixnextcell
		\node[coordinate] (m22) {}; \pgfmatrixnextcell
		\node[coordinate] (m23) {}; \pgfmatrixnextcell
		\node[dspsquare, scale=0.8] (m24) {$z^{-1}$}; \pgfmatrixnextcell
		\node[dspsquare, scale=0.8] (m25) {$z^{-1}$}; \pgfmatrixnextcell
		\node[coordinate] (m26) {}; \pgfmatrixnextcell
		\node[coordinate] (m27) {}; \pgfmatrixnextcell
		\node[coordinate] (m28) {}; \\
		\node[coordinate] (m31) {}; \pgfmatrixnextcell
		\node[dspnodefull] (m32) {}; \pgfmatrixnextcell
		\node[triangleleftright, label={above:$b_1$}, label={below:$\phantom{b_1}$}] (m33) {}; \pgfmatrixnextcell
		\node[dspadder] (m34) {}; \pgfmatrixnextcell
		\node[dspadder] (m35) {}; \pgfmatrixnextcell
		\node[trianglerightleft, label={above:$-a_1$}] (m36) {}; \pgfmatrixnextcell
		\node[dspnodefull] (m37) {}; \pgfmatrixnextcell
		\node[coordinate] (m38) {}; \\
		\node[coordinate] (m41) {}; \pgfmatrixnextcell
		\node[coordinate] (m42) {}; \pgfmatrixnextcell
		\node[coordinate] (m43) {}; \pgfmatrixnextcell
		\node[dspsquare, scale=0.8] (m44) {$z^{-1}$}; \pgfmatrixnextcell
		\node[dspsquare, scale=0.8] (m45) {$z^{-1}$}; \pgfmatrixnextcell
		\node[coordinate] (m46) {}; \pgfmatrixnextcell
		\node[coordinate] (m47) {}; \pgfmatrixnextcell
		\node[coordinate] (m48) {}; \\
		\node[coordinate] (m51) {}; \pgfmatrixnextcell
		\node[coordinate] (m52) {}; \pgfmatrixnextcell
		\node[triangleleftright, label={above:$b_2$}] (m53) {}; \pgfmatrixnextcell
		\node[coordinate] (m54) {}; \pgfmatrixnextcell
		\node[coordinate] (m55) {}; \pgfmatrixnextcell
		\node[trianglerightleft, label={above:$-a_2$}] (m56) {}; \pgfmatrixnextcell
		\node[coordinate] (m57) {}; \pgfmatrixnextcell
		\node[coordinate] (m58) {}; \\
	};
	
	\begin{scope}[start chain]
		\chainin (m11);
		\chainin (m12) [join=by dspline];
		\chainin (m13) [join=by dspline];
	\end{scope}
	\begin{scope}[start chain]
		\chainin (m12);
		\chainin (m22) [join=by dspline];
		\chainin (m32) [join=by dspline];
		\chainin (m33) [join=by dspline];
	\end{scope}
	\begin{scope}[start chain]
		\chainin (m32);
		\chainin (m42) [join=by dspline];
		\chainin (m52) [join=by dspline];
		\chainin (m53) [join=by dspline];
	\end{scope}
	\begin{scope}[start chain]
		\chainin (m53);
		\chainin (m54) [join=by dspline];
		\chainin (m44) [join=by dspline];
	\end{scope}
	\begin{scope}[start chain]
		\chainin (m44);
		\chainin (m34) [join=by dspconn];
	\end{scope}
	\begin{scope}[start chain]
		\chainin (m33);
		\chainin (m34) [join=by dspconn];
	\end{scope}
	\begin{scope}[start chain]
		\chainin (m34);
		\chainin (m24) [join=by dspline];
	\end{scope}
	\begin{scope}[start chain]
		\chainin (m24);
		\chainin (m14) [join=by dspconn];
	\end{scope}
	\begin{scope}[start chain]
		\chainin (m13);
		\chainin (m14) [join=by dspconn];
	\end{scope}
	\begin{scope}[start chain]
		\chainin (m14);
		\chainin (m15) [join=by dspconn];
	\end{scope}
	\begin{scope}[start chain]
		\chainin (m15);
		\chainin (m16) [join=by dspline];
		\chainin (m17) [join=by dspline];
		\chainin (m18) [join=by dspconn];
	\end{scope}
	\begin{scope}[start chain]
		\chainin (m17);
		\chainin (m27) [join=by dspline];
		\chainin (m37) [join=by dspline];
		\chainin (m36) [join=by dspline];
	\end{scope}
	\begin{scope}[start chain]
		\chainin (m37);
		\chainin (m47) [join=by dspline];
		\chainin (m57) [join=by dspline];
		\chainin (m56) [join=by dspline];
	\end{scope}
	\begin{scope}[start chain]
		\chainin (m56);
		\chainin (m55) [join=by dspline];
		\chainin (m45) [join=by dspline];
	\end{scope}
	\begin{scope}[start chain]
		\chainin (m45);
		\chainin (m35) [join=by dspconn];
	\end{scope}
	\begin{scope}[start chain]
		\chainin (m36);
		\chainin (m35) [join=by dspconn];
	\end{scope}
	\begin{scope}[start chain]
		\chainin (m35);
		\chainin (m25) [join=by dspline];
	\end{scope}
	\begin{scope}[start chain]
		\chainin (m25);
		\chainin (m15) [join=by dspconn];
	\end{scope}
	\end{tikzpicture}
\caption{Transposed Direct Form II for a second-order IIR filter.}
\label{fig:iirfilters}
\end{figure}

A large body of work
exists for the I step. Hardware filters involve multiplications with constants, for which optimization
techniques
have been extensively explored. In the multiplierless \emph{shift-and-add}-based methods, constant
multiplications are replaced by additions,
subtractions and bit shifts.
The associated optimization problem is known as the multiple constant multiplication (MCM) problem,
for which
heuristics~\cite{dm95a, vp07, KummZipfFaustChang_Pipelinedaddergraph_2012} as well as optimal approaches exist~\cite{acfm08, kfmzm13, Kumm_MultipleConstantMultiplication_2016_book, Kumm_OptimalConstantMultiplication_2018}.
Various cost functions are
possible: the high-level ones count the number of adders required to perform all multiplications, and
low-level ones counting the
number of full adders. This problem has been successfully modeled as (Mixed) Integer Linear
Programming (ILP) and solved using
efficient solvers such as CPLEX or Gurobi.
Another constant multiplication method, especially relevant for FPGAs, is based on precomputed tables
and is called Ken
Chapman multiplier (KCM), after its inventor~\cite{c94}.
It has also been successfully applied to digital filtering~\cite{fc11,kfmzm13,dffk19} and using this
method, an approach for
optimization of combined Q \& I steps has been proposed
\cite{VolkovaIstoanDeDinechinHilaire_TowardsHardwareIIR_2019}.

Both KCM and shift-and-add reduce arithmetic resources by sharing intermediate results.
However, this reduction strongly depends on the coefficient values, which are typically fixed in the
previous FD \& Q steps.
Hence, the obtained implementations are optimized only for one filter instance, or a small sub-set of the
overall design space, not
permitting overall optimal~solution.

\begin{figure}
	\centering
	\begin{tikzpicture}
		\begin{axis}[
			ytick pos=left,
			xtick pos=bottom,
			every tick label/.append style={},
			ymin=0,
			xmin=0,
			xmax=1,
			width=\linewidth,
			height=.7\linewidth,
			ylabel={Frequency response $\left|H\left(e^{i\omega}\right)\right|$},
			xlabel={Normalized frequency $\omega$},
			trig format plots=rad,
			legend style={},
		]

			\addplot [
				legend entry=Double precision,
				domain=0:1,
				samples=200,
				color=blue,
				 thick,
				dashed,
			]
			{sqrt((0.1705793014214943+0.08404437189569297*cos(2*pi*x)+0.12060492360624445*cos(pi*x)+0.12060492360624445*cos(pi*x))/(1.4642054606350485+0.6794017239917509*cos(2*pi*x)-0.40125490263735863*cos(pi*x)-1.181200719597322*cos(pi*x)))};
			\addplot [
				legend entry=Truncated to $5$ bits,
				domain=0:1,
				samples=200,
				color=red,
				thick,
				dotted,
			]
			{sqrt((0.16796875+0.0703125*cos(2*pi*x)+0.1171875*cos(pi*x)+0.1171875*cos(pi*x))/(1.53125+0.75*cos(2*pi*x)-1.25*cos(pi*x)-0.46875*cos(pi*x)))};
			\addplot [
				legend entry=Ours,
				domain=0:1,
				samples=200,
				color=green,
				 thick,
			]
			{sqrt((0.1494140625+0.0703125*cos(2*pi*x)+0.10546875*cos(pi*x)+0.10546875*cos(pi*x))/(1.61328125+0.75*cos(2*pi*x)-1.375*cos(pi*x)-0.515625*cos(pi*x)))};

			\addplot [domain=0:0.3,color=black,  thick] {1+0.1-0.01*4};
			\addplot [domain=0:0.3,color=black,  thick] {1-0.1+0.01*4};
			\addplot [domain=0.7:1,color=black,  thick] {0.1-0.01*4};
		\end{axis}
	\end{tikzpicture}
	\caption{An example of filter specification (\lp{1}{4} in our benchmarks), for which a standard FD \& Q approach fails to find 5-bit 
	coefficients, while our approach succeeds.}
	\label{fig:lp1x4_transferfunction}
\end{figure}
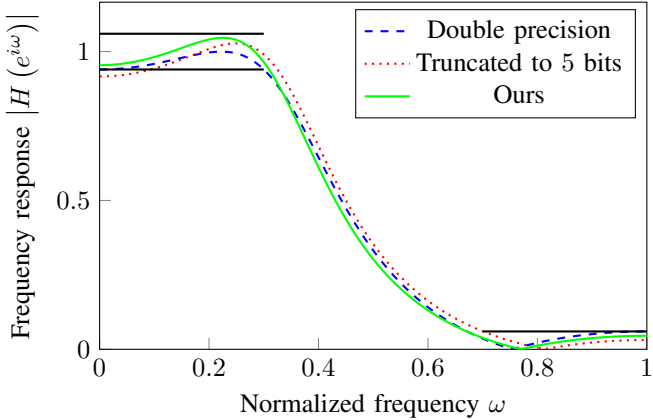

The most common model in formation of IIR filters is based on the decomposition of a higher-order filter
into a cascade of
second-order sections implemented with the Direct Form structure~\cite{Opp2009,
WangLiLi_Fixedpointdigital_2010,
	Leich_ToolboxdesignIIR_1997, JenkinsNayeri_Adaptivefiltersrealized_1986,
	DempsterMacleod_ComparisonIIRfilter_1995}.
We focus here on the optimal design of individual second-order filters (see \figurename~\ref{fig:iirfilters})
as a building brick for methods using cascaded forms, \eg{}, \cite{CuckooSearch}.
Our ambition is to solve the \textbf{combined FD \& Q \& I steps for a second-order IIR within one global
optimization.}
Even for one second-order filter this is a difficult non-linear combinatorial problem.
Recent work~\cite{CuckooSearch} attempts to solve the IIR filter design and cascading problem, while
targeting coefficients with
the minimal binary writing, but does not lead to optimal solutions due to restricted design-space and a
heuristic solver.

In contrast, in this paper we model the overall process as an ILP problem,
which is consequently
optimally solved using generic solvers.
With this work, the whole design space for the FD \& Q \& I problem is covered: we search for FxP
coefficients (\ie{}, the
quantization stage is implicit) that satisfy filter specifications while minimizing the cost of both structural
and multiplier blocks in a
multiplierless Transposed Direct Form II realization of a second-order IIR filter. We focus on the transposed form,
as it allows to model the multiplications of the input/output by filter coefficients as the MCM problem based on ILP.
\figurename~\ref{fig:lp1x4_transferfunction} exhibits an example of filter specifications, for which the
standard FD \& Q step does not find
$5$-bit coefficients satisfying the constraints, whilst our approach easily finds the coefficients that need only $8$ adders.

The proposed ILP formulation is building upon the optimal
MCM~\cite{Kumm_OptimalConstantMultiplication_2018} solution for
constant multiplications.
While different cost functions are possible, we start with a high-level metric counting number of adders,
leaving the low-level
metric of counting the full adders out of scope.

A similar approach has been recently proposed for optimizing the number of adders for FIR
filters~\cite{DesignOptimalMultiplierless_2019} but extending
its ideas
towards IIR filters requires overcoming several challenges:
expressing the filter design constraints for a rational transfer function in a linear way, guaranteeing the
stability and determining the
FxP format for multiplier blocks, which are described in Section~\ref{s:ILP-formulation}. Moreover,
we present a
design-space
reduction that significantly improves the solving time. As a result, we provide an automatic tool that,
given frequency specifications
and word length, within mere seconds determines the FxP filter coefficients that guarantee
the minimal number of adders for a
multiplierless implementation, as well as the multiplierless solution for each coefficient block. We
validate our approach on real-life and
artificial examples in Section~\ref{sec:expe} by automatically generating and synthesizing VHDL code on
FPGA targets.

\section{ILP Formulation for Second-Order IIR Filter Design}\label{s:ILP-formulation}

\subsection{Problem Definition}

Transposed Direct Form II structure of second order digital filters is represented in \figurename~\ref{fig:iirfilters}. Its
output is computed as
\begin{equation}\label{eq:CCDE}
y_n = \sum\limits_{k=0}^{2}{b_k u_{n-k}} - \sum\limits_{k=1}^{2}{a_k y_{n-k}},
\end{equation}
where $a_k, b_k \in \mathbb{R}$ are the filter coefficients. The corresponding transfer function is
\begin{equation}\label{eq:transferfunction}
H\!\left(z\right) = \frac{B\!\left(z\right)}{A\!\left(z\right)} = \frac{b_0 + b_1 z^{-1} + b_2 z^{-2}}{1 + a_1 z^{-1}
+ a_2 z^{-2}},
\end{equation}
with $z \in \mathbb{C}$. Designing a filter means finding the coefficients $a_k$ and $b_k$ such that the
filter is stable and given frequency constraints hold. These constraints can be expressed as bounds on
the filter frequency response
\begin{equation}\label{eq:specs}
\underline{\beta}\!\left(\omega\right) \leq \left|H\!\left(e^{j\omega}\right)\right| \leq
\overline{\beta}\!\left(\omega\right), \quad \forall \omega \in \closeint{0}{\pi},
\end{equation}
where $\underline{\beta}$ and $\overline{\beta}$ are the given frequency dependent lower and upper
bounds. They typically
encode constant bounds valid inside frequency intervals, but our definition~\eqref{eq:specs} allows to model them as functions of $\omega$.

There exists a large number of real-valued coefficient sets satisfying the frequency
specifications~\eqref{eq:specs} but only a subset
of those is representable in FxP arithmetic with a given word length.
In our approach, we directly search for $a_k$ and $b_k$ in a FxP
representation and introduce
their integer counterparts, $a'_k, b'_k \in \mathbb{Z}$, that are linked with the real-valued coefficients~via
\begin{align}
& a_k = 2^{-w+1} \, 2^{g_a} \, a'_k, && a'_k \in \closeint{-2^{w-1}}{2^{w-1}-1}, \label{al:a_int} \\
& b_k = 2^{-w+1} \, 2^{g_b} \, b'_k, && b'_k \in \closeint{-2^{w-1}}{2^{w-1}-1}, \label{al:b_int}
\end{align}
where $w$ is the word length (including sign bit), $g_a$ and $g_b$ are the largest Most Significant Bit
(MSB) positions of the FxP representations of $a_k$ and $b_k$, respectively.
It is typical to use the same word length for $a_k$ and $b_k$ but we allow different FxP formats in the
numerator and denominator of the transfer function.
The optimal design process proposed here is cast into the
integer programming framework by searching for the integer counterparts $a'_k$ and $b'_k$ within the
set of all representable
numbers defined by $w$, $g_b$ and $g_a$. While the word length $w$ is a user-given parameter in our
setting and is typically
desired to be as small as possible in order to save
resources\footnote{In Section~\ref{sec:hardware} we demonstrate that smaller coefficient word lengths
do not necessarily lead to
smaller resource consumption.}, the values of the MSB positions $g_b$ and $g_a$ are not known.
Hence, in order to provide an ILP
model one must first predict the precision and the range of the FxP coefficients.
This task is more challenging for IIR filters compared to FIR filters, where the MSB positions can be set to
$0$ and afterwards
controlled through the filter gain.
More precisely, the goal is to predict the MSB positions that, on the one hand, encompass all
representable coefficients, \ie{}, do
not lose solutions, and that maximize the precision, \ie{}, do not overestimate the MSB, on the other
hand.
We find upper bounds for $g_a$ and $g_b$ via a pre-processing routine described in Section~\ref{sec:stability_bounds}.
Further, we solve our model for a  range of lower MSBs with an external~loop.

As it has been successfully done for FIR filters~\cite{DesignOptimalMultiplierless_2019},
we unify the filter coefficients design and their quantization by searching directly for coefficients
satisfying the frequency constraints in their FxP representation: the overall objective is to find FxP
representations of filter coefficients such that filter specifications are met and the hardware
implementation is optimal with respect to the number of adders.
Extending this approach from FIR to IIR filters requires several technical contributions: the frequency
constraints~\eqref{eq:specs}
on the complex modulus of the rational transfer function~\eqref{eq:transferfunction} involve
highly nonlinear operations, which are linearized
in Section~\ref{sec:linearize}. We express stability constraints over the FxP coefficients and determine the
MSB positions for
coefficients $a_k$ and $b_k$ in Section~\ref{sec:stability_bounds}. Search space reductions are
presented in
Section~\ref{sec:searchspacereduction}. The connection of filter specification constraints with two
standard MCM ILP
models for its multiplierless implementation is presented in Section~\ref{sec:mcm}.
Finally, a short summary of the workflow is presented in
Section~\ref{ss:wrapping-up}.

\subsection{Formulating Linear Frequency Constraints}\label{sec:linearize}

In order to obtain an ILP model, we have to solve two issues related to the frequency
constraints~\eqref{eq:specs}. First of all,~\eqref{eq:specs} actually enforces one constraint for each
$\omega \in \closeint{0}{\pi}$. There are infinitely many such constraints to be satisfied. The
constraint~\eqref{eq:specs} is called semi-infinite (see~\cite{Hettich1993} for a survey on this class of
constraints). The standard discretization approach to handling semi-infinite constraints consists in
discretizing $\closeint{0}{\pi}$ into a finite set of frequencies $\Omega_d$, leading to a finite number of
constraints. This discretization can be dynamically updated, but we consider here a fixed discretization
and a \emph{a posteriori} verification
of the semi-infinite frequency constraint~\cite{VolkovaLauterHilaire_ReliableVerificationDigital_2017}. In
the rare cases where the verification fails, the verification procedure proposed
in~\cite{VolkovaLauterHilaire_ReliableVerificationDigital_2017} outputs a faulty frequency that can be
added to the discretized set of frequencies for a new trial.
Continuous variables generically lead to active
constraints, which require some theory and accurate algorithms with some adaptive discretization to
allow discovering active frequencies for optimal coefficients (see~\cite{Hettich1993,Marendet2020} for
details).
Integer variables generically do not lead to active constraints. This means that only a finite
number of faulty frequencies can happen after the ILP design, before we obtain coefficients verifying
rigorously the frequency specifications.
This reduction to a finite problem of the semi-infinite constraint is a positive side effect of searching
directly FxP representations of the~coefficients.

After this discretization of the semi-infinite constraint, a finite number of nonlinear constraints needs to
be handled. For a fixed
$\omega\in\Omega_d$, the constraint~\eqref{eq:specs} includes a complex absolute value, which
involves a square-root of the sum of
squared terms, and a fraction. By incorporating \eqref{eq:transferfunction} into \eqref{eq:specs}, then
multiplying with the
denominator $\left|A\!\left(e^{j\omega}\right)\right|$, and finally squaring the result we obtain a
constraint equivalent to
\eqref{eq:specs}
\begin{align}
& \left|A\!\left(e^{j\omega}\right)\right|^2 \underline{\beta}\!\left(\omega\right)^2 \leq
\left|B\!\left(e^{j\omega}\right)\right|^2 \leq \left|A\!\left(e^{j\omega}\right)\right|^2
\overline{\beta}\!\left(\omega\right)^2, \label{al:cst_transferfunction}
\end{align}
where,
\begin{align}
& \left|B\!\left(e^{j\omega}\right)\right|^2 = \sum\limits_{k=0}^{2}\sum\limits_{l=0}^{2}{b_k b_l
\cos\!\left(\left(k-l\right)\omega\right)}, \label{al:cst_transferfunction_b} \\
& \left|A\!\left(e^{j\omega}\right)\right|^2 = \sum\limits_{k=0}^{2}\sum\limits_{l=0}^{2}{a_k a_l
\cos\!\left(\left(k-l\right)\omega\right)}, \ \text{with } a_0 = 1. \label{al:cst_transferfunction_a}
\end{align}

Yet, in \eqref{al:cst_transferfunction_b} and \eqref{al:cst_transferfunction_a}, the filter coefficients are still
involved in bilinear terms $b_k b_l$ and $a_k a_l$. Billionnet \emph{et
al.}~\cite{BillionnetElloumiLambert_LinearReformulationsInteger_2008} proposed a method that allows
linearizing products of positive integers assuming that bounds on these numbers are known: consider
the product $z = xy$ where $x, y, z \in \mathbb{N}$, with $x \leq \overline{x}$ and $y \leq \overline{y}$.
The linearization~\cite{BillionnetElloumiLambert_LinearReformulationsInteger_2008} basically consists in
first rewriting one of the positive integers into its binary representation
\begin{equation}
\label{eq:inttobin}
x = \sum_{i=0}^{\left\lceil\log_2 \overline{x}\right\rceil+1}{2^i \, t_{x,i}},
\end{equation}
where $t_{x,i}$ are binary auxiliary variables. This constraint ensures that the bits $t_{x,i}$ encode the
value of $x$.
Then, the product $z=xy$ becomes a sum of products between the binary variables $t_{x,i}$ and the positive
integer $y$. Finally, such a binary by integer product is common and its well-known linearization involves
indicator or big~$M$ constraints~\cite{Glover_ImprovedLinearInteger_1975,
OralKettani_LinearizationProcedureQuadratic_1992}.

Here however, $x$ and $y$ correspond to filter coefficients that have no sign restriction. We extend the
linearization exposed above to signed integers by adding the auxiliary variables $x^+,y^+\in\mathbb{N}$
and $x^\sg,y^\sg\in\{0,1\}$, and link them by following constraints
\begin{align}
x^+ &= \left|x\right|, & y^+ &= \left|y\right|, \label{eq:lin_abs} \\
x^\sg &= \sign\!\left(x\right), & y^\sg &= \sign\!\left(y\right), \label{eq:lin_sign}
\end{align}
\begin{equation}
z^+ = x^+ y^+, \label{eq:lin_prodplus}
\end{equation}
where the linearization of the absolute values \eqref{eq:lin_abs} and the sign constraints
\eqref{eq:lin_sign} are well-known and involve indicator or big~$M$
constraints~\cite{BertsimasTsitsiklis_Introductionlinearoptimization_1997_book,
Mangasarian_Absolutevalueprogramming_2006}, and where $z^+ = x^+ y^+$ is the positive case we
already presented. Finally, $z = \pm z^+$ and the sign is determined by the values of $x^\sg$ and
$y^\sg$ directly in the model. This whole linearization relies on the fact that bounds on $x$ and $y$ are
known. This is addressed in the next subsection.

\subsection{Stability and Bounds on Filter Coefficients}\label{sec:stability_bounds}

Necessary and sufficient stability conditions for second-order filters are
well-known~\cite[Section~16.8]{Antoniou_DigitalFiltersAnalysis_2018_book} to be
\begin{align}
-2 < a_1 < 2, \label{eq:stability_a1} \\
\left|a_1\right|-1 < a_2 < 1. \label{eq:stability_a2}
\end{align}
As explained before, the absolute value is standardly linearized using indicator or big~$M$ constraints,
thus \eqref{eq:stability_a1}-\eqref{eq:stability_a2} actually fit in an ILP model. From these constraints it is
straightforward to derive bounds on $a_k$: $a_1 \in \openint{-2}{2}$ and $a_2 \in \openint{-1}{1}$. These
bounds are independent of the frequency specification of the filter and yield an upper bound, $g_a = 1$, for the MSB of the coefficients $a_k$.

Bounds on $b_k$, however, cannot be obtained independently of the filter specifications. Using the
bounds~\eqref{eq:stability_a1}-\eqref{eq:stability_a2} and the fact the cosine in
\eqref{al:cst_transferfunction_a} belongs to $\closeint{-1}{1}$, we deduce that
$\left|A\!\left(e^{j\omega}\right)\right|^2 \leq 16$. This bound together with the frequency specification
constraints \eqref{al:cst_transferfunction} lead to the constraint
\begin{equation}
\left|B\!\left(e^{j\omega}\right)\right|^2 \leq 16 {\overline{\beta}\!\left(\omega\right)}^2 \ ,
\label{eq:bounds_B}
\end{equation}
that needs to be satisfied by the coefficients $b_k$

As can be seen from~\eqref{al:cst_transferfunction_b}, $\left|B\!\left(e^{j\omega}\right)\right|^2$
is a quadratic form $b^TQb$ with respect to the variables $b_k$. Its characteristic matrix $Q$, whose
entries are $Q_{kl}=\cos((k-l)\omega)$, is symmetric and its spectrum is
$\{0,1-\cos(2\omega),2+\cos(2\omega)\}$. Its eigenvalues being non-negative, the inequality
constraint~\eqref{eq:bounds_B} is convex. As a consequence, some lower and upper bounds on the
coefficients $b_k$ can be computed by solving the convex quadratic problems consisting in minimizing
or maximizing $b_k$ subject to the convex quadratic constraints~\eqref{eq:bounds_B} for all frequencies
$\omega \in \Omega_{d'}$ where $\Omega_{d'}$ is a discretization of $\Omega$. The global
minimum and maximum of these problems are
required to be used as valid lower and upper bounds, such global extrema being easily computed by
local solvers since the quadratic
constraints are all convex and the cost is linear. In particular, common mixed ILP solvers can solve this kind
of nonlinear problem. In addition to allowing the linearization of frequency constraints, these bounds
permit to decide a first upper bound, $g_b$, for the MSB of the coefficients $b_k$.

These bounds do not fully take into account the specificity of the filter we are designing and use the worst case bounds 
in~\eqref{eq:bounds_B}. In any case, it is possible to compute
tighter bounds on the coefficients in order to reduce the search space as explained in Section~\ref{sec:searchspacereduction}.

\subsection{MCM for Direct-Form IIR Filters}\label{sec:mcm}

The FxP coefficients satisfying the above frequency and stability constraints are also subject to
hardware constraints.
We use the ILP-based hardware models for optimal multiplierless constant multiplication presented
in~\cite{Kumm_MultipleConstantMultiplication_2016_book,Kumm_OptimalConstantMultiplication_2018}
that can be readily
incorporated into our problem.
These models, given a set of constant coefficients, solve the minimization (\wrt{} number of adders) or
satisfiability (whether an
implementation with a given number of adders is feasible) problems. For the global IIR filter design
problem, the coefficients are,
however, the unknowns. Hence, the first task is designing a number of linking constraints that bind
the coefficient variables
for filter design with the inputs of an MCM problem. We achieve this by introducing a number of binary variables and refer to them as \textit{glue constraints} in
\figurename~\ref{fig:highlevel_iir}.

It should be noted, that two sets of MCM constraints must be designed, as there is one multiplier block
for coefficients $a_k$ and one for $b_k$, with the goal of minimization of the \textit{total} number of
adders, \ie{}, in both multiplier blocks and in the filter structure. The final cost
function for the adder minimization ILP formulation is then
\begin{align}
	\min A_{M_a} + A_{M_b} - \sum_{k=0}^2 \zeta_k^b - \sum_{k=1}^2 \zeta_k^a,
\end{align}
where $A_{M_a}$ and $A_{M_b}$ denote the number of adders in multiplier blocks for $a_k$ and $b_k$,
respectively; and
$\zeta^a_k$ and $\zeta^b_k$ are binary variables validating whether the respective filter coefficients are
zero. In other words, the
number of zero-valued coefficients is maximized, privileging sparsity in the implemented filters.
Moreover, such a formulation also
allows for FIR filter design as every $a_k$ can be equal to $0$.

\subsection{Search Space Reduction}\label{sec:searchspacereduction}
\subsubsection{Linearized Specifications Projections}
The bounds for $a_k$ and $b_k$ computed in the previous section allow implementing an ILP model by
fixing an FxP format and linearizing the bilinear terms involved in the frequency specifications. This first
incomplete ILP model does not include the geometry of the adder graph that represents the
multiplierless solution as defined in Section~\ref{sec:mcm}. Yet, in order to speed up the solving of the
complete ILP model, we tighten the bounds on the
coefficients by solving these incomplete ILPs that are simpler than the complete ILP model in the sense
that it does not include the geometry of the adder graph defined in Section~\ref{sec:mcm}: they consist
in minimizing or maximizing $b'_k$ subject to the stability constraints and the linearized frequency
constraints. These ILPs are obviously more difficult to solve than the continuous convex quadratic
problems used in the previous section to obtain crude bounds, but still much easier to solve than the
final complete model. As it has been shown in~\cite{DesignOptimalMultiplierless_2019} in the context of
the design of FIR filters, solving these simpler ILPs to obtain tighter bounds for the solving of the final ILP
is worthwhile.

\subsubsection{Symmetry Breaking}
The model contains some symmetries that can be broken in order to reduce the search space: the first
symmetry consists in simultaneously changing the sign of the values taken by $b_0$, $b_1$ and $b_2$,
the second symmetry consists in exchanging the values taken by $b_0$ and $b_2$. These two
symmetries leave the constraints on the coefficients $b_k$ unchanged, as can be easily seen on the
explicit expression
\begin{equation}\label{al:cst_transferfunction-explicit}
	b_0^2 + b_1^2 + b_2^2 + 2 b_0 b_1 \cos(\omega) + 2 b_1 b_2 \cos(\omega) + 2 b_0 b_2 \cos(2 \omega),
\end{equation}
of~\eqref{al:cst_transferfunction} involved in the constraint~\eqref{eq:bounds_B}: indeed,
\eqref{al:cst_transferfunction-explicit} is insensitive to changing all coefficients sign simultaneously and
to exchanging variables $b_0$ and $b_2$. Both symmetries have no incidence on the MCM problem defined in
Section~\ref{sec:mcm}, the existence of symmetric adder graphs being obvious for opposed or
exchanged coefficients.

As a consequence, from an arbitrary solution with
\begin{align}
	& b_0 = b^*_0,\ b_1 = b^*_1 \text{ and } b_2 = b^*_2, \label{al:initialsolution} \\
	\intertext{we can build three new solutions by simply applying these symmetries to obtain}
	& b_0 = -b^*_0,\ b_1 = -b^*_1 \text{ and } b_2 = -b^*_2, \\
	& b_0 = b^*_2,\ \phantom{-}b_1 = b^*_1\phantom{-} \text{ and } b_2 = b^*_0, \\
	& b_0 = -b^*_2,\ b_1 = -b^*_1 \text{ and } b_2 = -b^*_0. \label{al:bothsymmetries}
\end{align}
The fourth solution \eqref{al:bothsymmetries} is obtained by applying the two symmetries consecutively,
in any order. Note that some of these four symmetric solutions
\eqref{al:initialsolution}-\eqref{al:bothsymmetries} may be equal in some special cases, \eg{}, when
$b_0=b_1=b_2=0$.

Breaking these symmetries means finding additional constraints, called symmetry breaking constraints
(SBCs), that remove some symmetric solutions but keep at least one of
them~\cite{Walsh_GeneralSymmetryBreaking_2006}. If necessary, symmetric solutions that have not
been calculated due to the SBCs can be built afterward. In the best case, SBCs keep only one solution
among all symmetric solutions, in that case the SBCs are called total. Therefore, we expect here to
reduce the size of the search space by a ratio of four since solutions come within symmetry classes
containing four symmetric solutions. Finding SBCs for general symmetry groups is difficult, \eg{}, SBC
generation for symmetries consisting only of variable permutations rely on modern group theoretic
algorithms~\cite{GOLDSZTEJN2015105}. In our case, the symmetry group is generated by one variable
permutation and one central symmetry. Up to our knowledge, SBCs involving both variable permutations
and central symmetries have not yet been investigated and no SBC defined for them.

In order to derive these SBCs, the search space for $b_0$, $b_1$ and $b_2$, which is $\mathbb{R}^3$, is divided into four areas:
\begin{align}
	& \Sigma_1 = \left\{\left(b_0, b_1, b_2\right) \in \mathbb{R}^3\ |\ b_0 \geq \left|b_2\right|\right\}, \\
	& \Sigma_2 = \left\{\left(b_0, b_1, b_2\right) \in \mathbb{R}^3\ |\ -b_0 \geq \left|b_2\right|\right\}, \\
	& \Sigma_3 = \left\{\left(b_0, b_1, b_2\right) \in \mathbb{R}^3\ |\ b_2 \geq \left|b_0\right|\right\}, \\
	& \Sigma_4 = \left\{\left(b_0, b_1, b_2\right) \in \mathbb{R}^3\ |\ -b_2 \geq \left|b_0\right|\right\}.
\end{align}
Then, one can verify that: a solution lying inside $\Sigma_2$ moves to $\Sigma_1$ applying the sign symmetry;
a solution lying inside $\Sigma_3$ moves to $\Sigma_1$ applying the exchange symmetry; a solution
lying inside $\Sigma_4$ moves to $\Sigma_1$ applying both symmetries consecutively. As a
consequence, one can restrict the search to $\Sigma_1$ and reconstruct all solutions using symmetries.
This restriction to $\Sigma_1$ is achieved by adding the SBC
\begin{equation}\label{eq:symbreak}
	b_0 \geq \left|b_2\right|
\end{equation}
to the model. 
Restricting to another $\Sigma_i$ would be lead to another SBC, with an equivalent improvement of the
resolution~process.

For completeness, it should be noted that on the frontier $\Sigma_k \cap \Sigma_l$ between areas, two equivalent solutions
might still be kept despite the SBC. However, resolving this issue is counterproductive, because
the great majority of the search space lies in the interior of the sets $\Sigma_i$, and removing the
symmetries on the boundaries would introduce many additional constraints.

\subsection{Wrapping Up}\label{ss:wrapping-up}

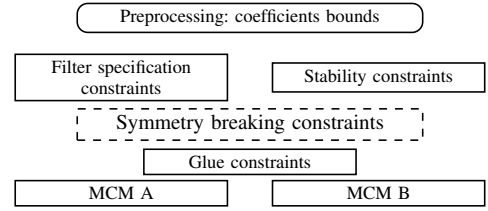
\begin{figure}
	\centering\scriptsize
	\begin{tikzpicture}[node distance=2mm, auto]
	\tikzset{mynode/.style = {
			shape=rectangle,
			draw=black, semithick,
			text width=0.3\linewidth,
			align=center}
	}
	\tikzset{mynotsosurenode/.style = {
			shape=rectangle,
			draw=black, semithick, dashed,
			text width=0.28\linewidth,
			align=center,
			font=\footnotesize}
	}
	\node[mynode] (glue) [draw] {Glue constraints};
	\node (activateornot) [above=1.5mm of glue] {\phantom{.}};
	\node (constraints) [above=5mm of activateornot] {};
	\node[mynode, rounded corners, text width=0.5\linewidth] (preprocess) [above=5mm of constraints] {Preprocessing: coefficients bounds};
	\node[mynode] (specifications) [left=of constraints] {Filter specification constraints};
	\node[mynode] (stability) [right=of constraints] {Stability constraints};
	\node (mcm) [below=1.5mm of glue] {};
	\node[mynode] (mcm1) [left=2mm of mcm] {MCM A};
	\node[mynode] (mcm2) [right=2mm of mcm] {MCM B};
	\node[mynotsosurenode, text width=0.5\linewidth] (symmetry) [above=1mm of glue] {Symmetry breaking constraints};
	\end{tikzpicture}
	\caption{High-level structure of the global ILP model IIRoptim.}
	\label{fig:highlevel_iir}
\end{figure}

To wrap it up, our approach is to model both the filter design and the design of a constant multiplication
scheme through
shift-and-add using a global ILP problem. On top of that, we search for coefficients directly as integers,
\ie{}, in a FxP format
with user-given word length. Since filter coefficients, depending on the filter specification and word
length, can have different MSB
positions, we first perform a pre-processing. This pre-processing consists in solving a quadratic convex
optimization problem
yielding tight and rigorous bounds on filter coefficients and permitting to define coherent MSB positions and
construct the design space
for the main model. As \figurename~\ref{fig:highlevel_iir} shows, the high-level IIR model consists of the
following constraints:
\begin{itemize}
\setlength\itemsep{0em}
	\item linearized frequency-specification constraints~\eqref{al:cst_transferfunction}-\eqref{al:cst_transferfunction_a}
	;
	\item stability constraints~\eqref{eq:stability_a1}-\eqref{eq:stability_a2};
	\item (optional) symmetry breaking constraints~\eqref{eq:symbreak} enabling a significant reduction of the design space;
	\item constraints responsible for the design of optimal constant-multiplication blocks for $a_k$ and
	$b_k$ using~\cite{Kumm_MultipleConstantMultiplication_2016_book} or~\cite{Kumm_OptimalConstantMultiplication_2018};
	\item so-called glue constraints, connecting the unknown filter coefficients with multiplier blocks analogously to~\cite{DesignOptimalMultiplierless_2019}.
\end{itemize}

The above ILP model can now be solved using any available ILP solver. Moreover, despite its
non-linearity, the pre-processing
problem can be solved using most generic ILP solvers as well, thanks to its convexity.

%


\section{Experimental Results and Discussion}\label{sec:expe}



\subsection{Implemented Toolflow}

\begin{figure}
\definecolor{backgroundcolor}{RGB}{202, 231, 193}
\definecolor{whiteorblack}{RGB}{0,0,0}
\begin{tikzpicture}[font=\sffamily]
	\tikzset{basicnode/.style = {
	align=center
	}
	}
	\tikzset{smallnodelabel/.style = {
		prefix after command= {\pgfextra{\tikzset{every label/.style={font=\scriptsize}}}}}
	}
	\tikzset{myfitnode/.style = {
		shape=rectangle,
		rounded corners,
		semithick, draw=black}
	}
	\tikzset{mysimplenode/.style = {
		shape=rectangle,
		draw=black, semithick,
		text centered,
		align=center,
		minimum height=3em,
		font=\small}
	}
	\tikzset{mynotsosurenode/.style = {
		shape=rectangle,
		draw=black, semithick, dashed,
		align=left,
		font=\small}
	}
	\node[basicnode] (centernode) {};
	\node[basicnode] (lownode) [below=1mm of centernode] {};
	\node[basicnode, text=whiteorblack] (jiir2hw) [above=1mm of centernode] {Optimal order-2 IIR
	hardware\\generator};
	\node[mysimplenode, draw=whiteorblack, text=whiteorblack] (jiir2ag) [left=6mm of lownode] {IIRoptim};
	\node[mysimplenode, draw=whiteorblack, text=whiteorblack] (flopoco) [right=6mm of lownode] {FloPoCo};

	\draw[->, shorten >= 1pt, shorten <= 1pt, color=whiteorblack] (jiir2ag) -- node [midway, above] {\scriptsize coeffs} node [midway, below] {\scriptsize addergraphs} (flopoco);

	\begin{scope}[on background layer]
		\node[myfitnode, font=\scriptsize, text centered, draw=whiteorblack, fill=backgroundcolor, fit={(jiir2hw) (jiir2ag) (flopoco)}] (wholeblock) {};
	\end{scope}

	\node[mysimplenode] (vhdl) [right=3mm of wholeblock] {.vhdl};
	\node[mynotsosurenode] (filterspecs) [left=3mm of wholeblock] {Frequency\phantom{..}\\specifications\phantom{..}
	\\\vspace{-0.5em}\\ {\hfill\footnotesize \textcolor{blue}{filter spec.}}};
	\node[mynotsosurenode] (hardwarespecs) [below=3mm of wholeblock] {\phantom{..}input format \quad output format \quad
	FPGA freq\phantom{..} \\\vspace{-0.5em}\\ {\hfill\footnotesize \textcolor{blue}{hardware spec.}}};

	\draw[->, semithick, shorten >= 1pt, shorten <= 1pt] (wholeblock) -- (vhdl);
	\draw[->, semithick, shorten >= 1pt, shorten <= 1pt] (filterspecs) -- (wholeblock);
	\draw[->, semithick, shorten >= 1pt, shorten <= 1pt] (hardwarespecs) -- (wholeblock);


\end{tikzpicture}
\caption{Interface of the proposed tool. }
\label{fig:tool}
\end{figure}
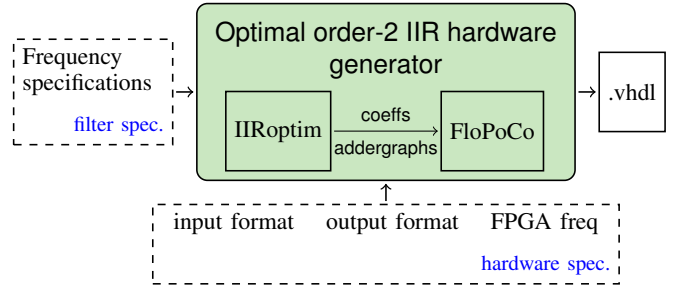

We implemented our approach into a tool, whose minimalist interface is shown in \figurename~\ref{fig:tool}. 
The input specification includes the
frequency-domain
specification of the filter, and also the information for hardware implementation, \ie{}, the coefficient word
length, the input/output
formats and performance parameters, such as the required frequency for target technology (currently different FPGA targets).
All the other parameters (filter coefficients and their FxP format, the multiplierless operator architecture, the representation of intermediate data, etc.)
are determined automatically as a part of a global optimization process. One of the goals of our tool is to bring the
attention of the filter designer to the higher-level parameters, \eg{}, filter specifications, while relying on
our optimal implementations.

Our new ILP model and the front-end of the proposed tool are implemented in \verb|julia| language,
which offers a unified access
to the major ILP solvers through the \verb|JuMP| library.
Given the user input, we first construct an ILP model and solve it with one of many open-access or
commercial solvers available
through \verb|JuMP|, such as Gurobi, CPLEX, GLPK, etc. The result of the global optimization problem,
\ie{}, the list of filter
coefficients and the adder graphs defining the optimal shift-and-add architectures, is then passed on to
the FloPoCo\footnote{\url{http://flopoco.org/} \label{flopoco}} hardware
code generator \cite{DinechinPasca_DesigningCustomArithmetic_2011}.
FloPoCo is the state-of-the-art tool for the design and automatic generation of fixed/floating-point
arithmetic
cores. We implemented a new operator \texttt{FixIIRShiftAdd}, generating faithfully-rounded multiplierless IIR filters, \ie{}, only the last output bit might be erroneous and all other bits are guaranteed to be correct. The new operator alleviates the filter designer from all internal architectural decisions and presents a final VHDL code.

The tool and all benchmarks, are freely available\footref{flopoco}\,\footnote{\url{https://gitlab.univ-nantes.fr/volkova-a/jiir2hw}} and reproducible.

\subsection{Set of Benchmarks and Comparison Approaches} \label{sec:benchmark}

\begin{table}[t]
	\centering
	\caption{Sets of lowpass filters used for the IIR experiments. First the set with decreasing $\delta$, next the sets with increasing passband and stopband. Finally, a lowpass filter.}
	\setlength{\tabcolsep}{5pt}
	{\scriptsize
	\begin{tabular}{l|cccc}\toprule\label{tab:benchmarks_set}
		Benchmarks & \lp{1}{k} & \lp{2}{k} & \lp{3}{k} & \filtername{lp4} \\
		\midrule
		$k$ & $\left\{0, 1, \dots, 6\right\}$ & $\left\{0, 1, \dots, 4\right\}$ & $\left\{0, 1, \dots, 4\right\}$ & $-$ \\
		passband/$\pi$ & $\closeint{0}{0.3}$ & $\closeint{0}{0.3 + 0.05k}$ & $\closeint{0}{0.3}$ & $\closeint{0}{0.5}$ \\
		stopband/$\pi$ & $\closeint{0.7}{1}$ & $\closeint{0.7}{1}$ & $\closeint{0.7 - 0.05k}{1}$ & ${0.9}{1}$ \\
		$\delta$ & $0.1 - 0.01k$ & $0.1$ & $0.1$ & $0.1$ \\
		\bottomrule
	\end{tabular}
	}
\end{table}

\paragraph*{Benchmarks}Although the design of second-order IIR filters is an important part of the
design of larger order filters, benchmarks are rarely targeting frequency specifications of individual
second-order sections.
Hence, we use three sets of filter specifications with increasing filter design difficulty that could without
doubt be used in real-life applications. In addition to that, we add another artificial low-pass filter, and a
real-life example from~\cite{ VolkovaIstoanDeDinechinHilaire_TowardsHardwareIIR_2019}.

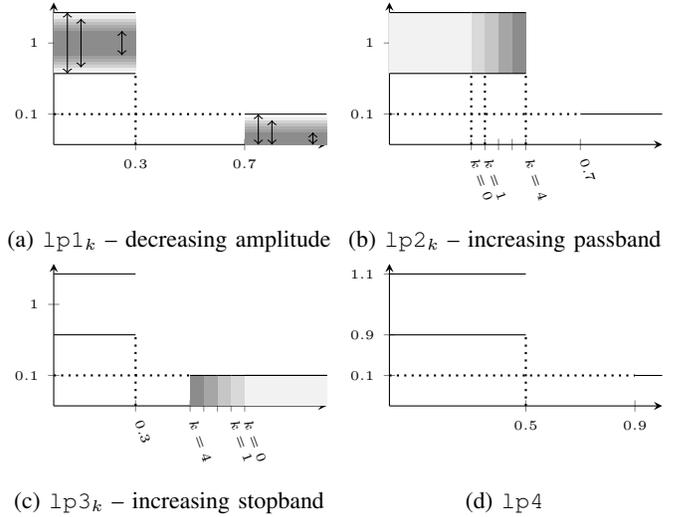
\begin{figure}
	\newcommand\scalebands{3}
	\newcommand\passbandWidth{0.1*\scalebands}
	\newcommand\stopbandWidth{0.1*\scalebands}
	\newcommand\bandWidthStep{0.01*\scalebands}
	\centering
	\begin{subfigure}[b]{0.49\linewidth}
		\centering
		\begin{tikzpicture}
			\draw [draw=white] (0,-0.9) rectangle (3.7,2);
			\begin{axis}[
				ytick pos=left,
				xtick pos=bottom,
				every tick label/.append style={font=\tiny},
				ymin=0,
				ymax=1+\passbandWidth+0.1,
				xmin=0,
				xmax=1,
				width=1.2\linewidth,
				height=0.8\linewidth,
				xtick={0.3,0.7},
				xticklabels={$0.3$, $0.7$},
				ytick={\stopbandWidth, 1},
				yticklabels={$0.1$, $1$},
				axis x line=bottom,
				axis y line=left,
				trig format plots=rad,
				legend style={font=\tiny},
			]
				\fill[gray!5!] (0,1-\passbandWidth+\bandWidthStep*0) rectangle (0.3,1+\passbandWidth-\bandWidthStep*0);
				\fill[gray!15!] (0,1-\passbandWidth+\bandWidthStep*1) rectangle (0.3,1+\passbandWidth-\bandWidthStep*1);
				\fill[gray!35!] (0,1-\passbandWidth+\bandWidthStep*2) rectangle (0.3,1+\passbandWidth-\bandWidthStep*2);
				\fill[gray!50!] (0,1-\passbandWidth+\bandWidthStep*3) rectangle (0.3,1+\passbandWidth-\bandWidthStep*3);
				\fill[gray!65!] (0,1-\passbandWidth+\bandWidthStep*4) rectangle (0.3,1+\passbandWidth-\bandWidthStep*4);
				\fill[gray!75!] (0,1-\passbandWidth+\bandWidthStep*5) rectangle (0.3,1+\passbandWidth-\bandWidthStep*5);
				\fill[gray!90!] (0,1-\passbandWidth+\bandWidthStep*6) rectangle (0.3,1+\passbandWidth-\bandWidthStep*6);

				\fill[gray!5!] (0.7,0) rectangle (1,\stopbandWidth-\bandWidthStep*0);
				\fill[gray!15!] (0.7,0) rectangle (1,\stopbandWidth-\bandWidthStep*1);
				\fill[gray!35!] (0.7,0) rectangle (1,\stopbandWidth-\bandWidthStep*2);
				\fill[gray!50!] (0.7,0) rectangle (1,\stopbandWidth-\bandWidthStep*3);
				\fill[gray!65!] (0.7,0) rectangle (1,\stopbandWidth-\bandWidthStep*4);
				\fill[gray!75!] (0.7,0) rectangle (1,\stopbandWidth-\bandWidthStep*5);
				\fill[gray!90!] (0.7,0) rectangle (1,\stopbandWidth-\bandWidthStep*6);

				\addplot [domain=0:0.3,color=black] {1+\passbandWidth};
				\addplot [domain=0:0.3,color=black] {1-\passbandWidth};
				\addplot [domain=0.7:1,color=black] {\stopbandWidth};

				\draw[<->] (0.05, 1-\passbandWidth+\bandWidthStep*0) -- (0.05, 1+\passbandWidth-\bandWidthStep*0);
				\draw[<->] (0.1, 1-\passbandWidth+\bandWidthStep*2) -- (0.1, 1+\passbandWidth-\bandWidthStep*2);
				\draw[<->] (0.25, 1-\passbandWidth+\bandWidthStep*6) -- (0.25, 1+\passbandWidth-\bandWidthStep*6);

				\draw[<->] (0.7+0.05, 0) -- (0.7+0.05, \stopbandWidth-\bandWidthStep*0);
				\draw[<->] (0.7+0.1, 0) -- (0.7+0.1, \stopbandWidth-\bandWidthStep*2);
				\draw[<->] (0.7+0.25, 0) -- (0.7+0.25, \stopbandWidth-\bandWidthStep*6);

				\addplot [domain=0:0.7,thick, dotted, color=black] {\stopbandWidth};

				\addplot [thick, dotted, color=black] coordinates {(0.3,0)(0.3,1-\passbandWidth)};
			\end{axis}
		\end{tikzpicture}
		\caption{\lp{1}{k} -- decreasing amplitude}
	\end{subfigure}
	\begin{subfigure}[b]{0.49\linewidth}
		\centering
		\begin{tikzpicture}
		\draw [draw=white] (0,-0.9) rectangle (3.7,2);
		\begin{axis}[
		ytick pos=left,
		xtick pos=bottom,
		every tick label/.append style={font=\tiny},
		ymin=0,
		ymax=1+\passbandWidth+0.1,
		xmin=0,
		xmax=1,
		xtick={0.3,0.35,0.4,0.45,0.5,0.7},
		xticklabels={$k=0$, $k=1$,,, $k=4$,$0.7$},
		width=1.2\linewidth,
		height=0.8\linewidth,
		ytick={\stopbandWidth, 1},
		yticklabels={$0.1$, $1$},
		axis x line=bottom,
		axis y line=left,
		trig format plots=rad,
		x tick label style={rotate=-70,anchor=west},
		legend style={font=\tiny},
		]
		\fill[gray!10!] (0,1-\passbandWidth) rectangle (0.3,1+\passbandWidth);
		\fill[gray!30!] (0.3,1-\passbandWidth) rectangle (0.35,1+\passbandWidth);
		\fill[gray!45!] (0.35,1-\passbandWidth) rectangle (0.4,1+\passbandWidth);
		\fill[gray!70!] (0.4,1-\passbandWidth) rectangle (0.45,1+\passbandWidth);
		\fill[gray!90!] (0.45,1-\passbandWidth) rectangle (0.5,1+\passbandWidth);
		\addplot[domain=0:0.3,color=black] {1+\passbandWidth};
		\addplot[domain=0:0.3,color=black] {1-\passbandWidth};
		\addplot[domain=0.3:0.35,color=black] {1+\passbandWidth};
		\addplot[domain=0.3:0.35,color=black] {1-\passbandWidth};
		\addplot[domain=0.35:0.4,color=black] {1+\passbandWidth};
		\addplot[domain=0.35:0.4,color=black] {1-\passbandWidth};
		\addplot[domain=0.4:0.45,color=black] {1+\passbandWidth};
		\addplot[domain=0.4:0.45,color=black] {1-\passbandWidth};
		\addplot[domain=0.45:0.5,color=black] {1+\passbandWidth};
		\addplot[domain=0.45:0.5,color=black] {1-\passbandWidth};

		\addplot [domain=0.7:1, color=black] {\stopbandWidth};

		\addplot [domain=0:0.7, thick, dotted, color=black] {\stopbandWidth};

		\addplot [thick, dotted, color=black] coordinates {(0.3,0)(0.3,1-\passbandWidth)};
		\addplot [thick, dotted, color=black] coordinates {(0.35,0)(0.35,1-\passbandWidth)};
		\addplot [thick, dotted, color=black] coordinates {(0.5,0)(0.5,1-\passbandWidth)};
		\end{axis}
		\end{tikzpicture}
		\caption{\lp{2}{k} -- increasing passband}
	\end{subfigure}
	\begin{subfigure}[b]{0.49\linewidth}
		\centering
		\begin{tikzpicture}
		\draw [draw=white] (0,-0.9) rectangle (3.7,2);
		\begin{axis}[
		ytick pos=left,
		xtick pos=bottom,
		every tick label/.append style={font=\tiny},
		ymin=0,
		ymax=1+\passbandWidth+0.1,
		xmin=0,
		xmax=1,
		width=1.2\linewidth,
		height=0.8\linewidth,
		xtick={0.3,0.5,0.55,0.6,0.65,0.7},
		xticklabels={$0.3$,$k=4$,,, $k=1$, $k=0$},
		axis x line=bottom,
		axis y line=left,
		ytick={\stopbandWidth, 1},
		yticklabels={$0.1$, $1$},
		trig format plots=rad,
		x tick label style={rotate=-70,anchor=west},
		legend style={font=\tiny},
		]
		\addplot [domain=0:0.3,color=black] {1+\passbandWidth};
		\addplot [domain=0:0.3,color=black] {1-\passbandWidth};

		\fill[gray!10!] (0.7,0) rectangle (1,\passbandWidth);
		\fill[gray!30!] (0.65,0) rectangle (0.7,\passbandWidth);
		\fill[gray!45!] (0.6,0) rectangle (0.65,\passbandWidth);
		\fill[gray!70!] (0.55,0) rectangle (0.6,\passbandWidth);
		\fill[gray!90!] (0.5,0) rectangle (0.55,\passbandWidth);

		\addplot [domain=0.7:1, color=black] {\stopbandWidth};
		\addplot [domain=0.65:7, color=black] {\stopbandWidth};
		\addplot [domain=0.6:0.65, color=black] {\stopbandWidth};
		\addplot [domain=0.55:0.6, color=black] {\stopbandWidth};
		\addplot [domain=0.5:0.55, color=black] {\stopbandWidth};

		\addplot [domain=0:0.5, thick, dotted, color=black] {\stopbandWidth};

		\addplot [thick, dotted, color=black] coordinates {(0.3,0)(0.3,1-\passbandWidth)};
		\end{axis}
		\end{tikzpicture}
		\caption{\lp{3}{k} -- increasing stopband}
	\end{subfigure}
	\begin{subfigure}[b]{0.49\linewidth}
		\centering
		\begin{tikzpicture}
		\draw [draw=white] (0,-0.9) rectangle (3.7,2);
		\begin{axis}[
		ytick pos=left,
		xtick pos=bottom,
		every tick label/.append style={font=\tiny},
		ymin=0,
		ymax=1+\passbandWidth+0.1,
		xmin=0,
		xmax=1,
		width=1.2\linewidth,
		height=0.8\linewidth,
		xtick={0.5,0.9},
		ytick={\passbandWidth, 1-\passbandWidth, 1+\passbandWidth},
		yticklabels={$0.1$, $0.9$, $1.1$},
		axis x line=bottom,
		axis y line=left,
		trig format plots=rad,
		legend style={font=\tiny},
		]
		\addplot [domain=0:0.5, color=black] {1+0.1*\scalebands};
		\addplot [domain=0:0.5, color=black] {1-0.1*\scalebands};
		\addplot [domain=0.9:1, color=black] {0.1*\scalebands};
		
		\addplot [domain=0:0.9, thick, dotted, color=black] {\stopbandWidth};
		
		\addplot [thick, dotted, color=black] coordinates {(0.5,0)(0.5,1-\stopbandWidth)};
		\end{axis}
		\end{tikzpicture}
		\caption{\filtername{lp4}}
	\end{subfigure}
	\caption{Proposed families of benchmarks.}
	\label{fig:bench_specs}
\end{figure}

The normalized low-pass filter specifications are here defined as
\begin{align*}
1 - \delta \leq &	\left|H(e^{i\omega})\right| \leq 1+\delta, \quad &\forall \omega \in
\closeint{0}{\omega_p},
\; &\text{ (passband)} \\
0 \leq &	\left|H(e^{i\omega})\right| \leq \delta, \quad &\forall \omega \in \closeint{\omega_s}{1}. \;
&\text{ (stopband)}
\end{align*}
We fix the initial passband to $\closeint{0}{0.3}$, stopband to $\closeint{0.7}{1}$ and $\delta=0.1$. Then, for
each of the
families of filter specifications, we vary one of the parameters in dependence of a variable $k$ to increase the filter design difficulty.
The detailed frequency specifications for each family of filters are given in
Table~\ref{tab:benchmarks_set}, and their graphical representation is sketched in
\figurename~\ref{fig:bench_specs}. For example, in family \filtername{lp1}, the $\delta$ varies from $0.1$
to $0.04$ with
step $0.01k$ where $k=0,\ldots,6$. However, the designs were possible only up to $k=5$, reaching the
maxium design possibilities for second-order IIR filters. Analogously, the families \filtername{lp2} and \filtername{lp3}
increase/reduce the pass/stopband, respectively. The filter specification \filtername{lp4} is a lowpass with a short stopband.

\begin{figure}
	\centering
	\begin{tikzpicture}
	\begin{axis}[
	ytick pos=left,
	xtick pos=bottom,
	legend style={at={(0.9,0.1)}, anchor=south east},
	ymin=0,
	xmin=-0.05,
	xmax=1,
	width=\linewidth,
	height=.5\linewidth,
	ylabel={$\left|H\left(e^{i\omega}\right)\right|$},
	xlabel={Normalized frequency $\omega$},
	]
	\addplot [
	color=black,
	 thick,
	] coordinates
	{(0.0, 0.01997896950578355) (0.01, 0.7101620632256287) (0.02, 0.9039096080179476) (0.03, 0.9609505189078075) (0.04, 0.9836379338588604) (0.05, 0.994700396881582) (0.06, 1.0008690497573036) (0.07, 1.0046447327141095) (0.08, 1.0071183799521026) (0.09, 1.0088249358402237) (0.1, 1.0100509711015229) (0.11, 1.0109609795665886) (0.12, 1.0116547571935135) (0.13, 1.012195657854909) (0.14, 1.0126254510373578) (0.15, 1.0129725715336149) (0.16, 1.0132569152183268) (0.17, 1.0134927385113728) (0.18, 1.0136904718329536) (0.19, 1.01385788791649) (0.2, 1.014000874386352) (0.21, 1.014123956614239) (0.22, 1.014230658971093) (0.23, 1.0143237591105754) (0.24, 1.0144054699936893) (0.25, 1.014477572193959) (0.26, 1.014541511413719) (0.27, 1.014598471282735) (0.28, 1.0146494283462684) (0.29, 1.0146951940524276) (0.3, 1.014736447135597) (0.31, 1.0147737588262111) (0.32, 1.0148076126467267) (0.33, 1.0148384200825373) (0.34, 1.0148665330815216) (0.35, 1.0148922540948997) (0.36, 1.0149158441968604) (0.37, 1.0149375296918048) (0.38, 1.0149575075227262) (0.39, 1.0149759497229867) (0.4, 1.0149930071000521) (0.41, 1.0150088122989465) (0.42, 1.0150234823619473) (0.43, 1.0150371208770121) (0.44, 1.0150498197887345) (0.45, 1.0150616609310645) (0.46, 1.0150727173295888) (0.47, 1.0150830543121194) (0.48, 1.0150927304591606) (0.49, 1.0151017984201003) (0.5, 1.015110305616359) (0.51, 1.0151182948490423) (0.52, 1.0151258048256158) (0.53, 1.0151328706177094) (0.54, 1.0151395240601369) (0.55, 1.0151457940995923) (0.56, 1.0151517071001486) (0.57, 1.0151572871115495) (0.58, 1.0151625561053792) (0.59, 1.0151675341834219) (0.6, 1.0151722397618879) (0.61, 1.0151766897346253) (0.62, 1.0151808996180218) (0.63, 1.015184883679879) (0.64, 1.0151886550542555) (0.65, 1.0151922258439823) (0.66, 1.0151956072123296) (0.67, 1.0151988094651045) (0.68, 1.0152018421242945) (0.69, 1.0152047139942253) (0.7, 1.015207433221076) (0.71, 1.0152100073464863) (0.72, 1.0152124433559093) (0.73, 1.0152147477222722) (0.74, 1.0152169264454407) (0.75, 1.0152189850879305) (0.76, 1.0152209288072471) (0.77, 1.0152227623851995) (0.78, 1.0152244902544822) (0.79, 1.0152261165228047) (0.8, 1.0152276449947872) (0.81, 1.0152290791918521) (0.82, 1.0152304223702768) (0.83, 1.0152316775375951) (0.84, 1.015232847467474) (0.85, 1.0152339347132104) (0.86, 1.0152349416199584) (0.87, 1.015235870335791) (0.88, 1.015236722821692) (0.89, 1.0152375008605523) (0.9, 1.0152382060652494) (0.91, 1.0152388398858745) (0.92, 1.015239403616158) (0.93, 1.0152398983991473) (0.94, 1.01524032523218) (0.95, 1.015240684971187) (0.96, 1.0152409783343637) (0.97, 1.0152412059052285) (0.98, 1.0152413681350962) (0.99, 1.0152414653449857) (1.0, 1.0152414977269706) };
	\end{axis}
	\end{tikzpicture}
	\caption{Frequency response of the compensator  \filtername{hp0}~\cite{CarlettaVeilletteKrachFang_Determiningappropriateprecisions_2003, 
	SarbisheiRadeckaZilic_AnalyticalOptimizationBit_2012}.}
	\label{fig:hp0}
\end{figure}
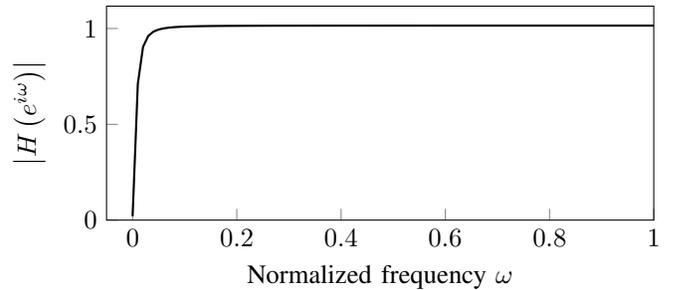

Finally, our last benchmark \filtername{hp0} is a highpass filter (\figurename~\ref{fig:hp0}), which is a compensator used in a
magnetic-bearing
control system and was derived by discretizing the analog
controller~\cite{KrachFrackeltonCarlettaVeillette_FPGAbasedimplementation_2003,
SarbisheiRadeckaZilic_AnalyticalOptimizationBit_2012}. The recent
result~\cite{VolkovaIstoanDeDinechinHilaire_TowardsHardwareIIR_2019} uses this filter to demonstrate a
KCM-based faithfully-rounded implementation of IIR filters, hence permits a direct comparison.
Even though this filter is not defined in terms of frequency specifications but by its frequency response
(sole poles and zeroes are given in literature), the versatility of an ILP modeling permits to easily
integrate frequency response bounds as functions of $\omega$ and not simply constants.


\paragraph*{Comparison approaches}
We aim at comparing with the classical and state-of-the-art approaches for the IIR design comparing high level metrics like adder counts and implementation results for FPGAs.
The classical approach passes through three steps
\begin{enumerate}[align=left,leftmargin=0mm,labelwidth=-9mm]
	\item \textbf{FD:} in our case double-precision filter coefficients are obtained, if possible, with
	Matlab's elliptic
	method;
	\item \textbf{Q:} as in Matlab's Fixed-Point design Toolbox, we convert the double coefficients to
	FxP
	with the user-specified word length and round-to-nearest mode. Post-design quantization often
	leads to errors in frequency response or instabilities, hence we increase the word length \st{} the
	frequency-domain error is below a threshold ($<10^{-7}$) and the filter is stable;
	\item \textbf{I:} generic multipliers provided by hardware manufacturers are often used. By default, the
	digital
	signal processing (DSP)
	blocks (whose availability and number on each FPGA board vary), are allowed but it might be
	interesting to disable them to provide comparison with purely
	LUT-based implementations. We also compare with the state-of-the-art constant multiplications based
	on KCM~\cite{VolkovaIstoanDeDinechinHilaire_TowardsHardwareIIR_2019} and
	MCM~\cite{Kumm_OptimalConstantMultiplication_2018} methods.
\end{enumerate}

Different combinations of the above steps are possible. Each of the benchmark filter specifications will
be implemented using the following approaches:
\begin{itemize}
	\item 3-step Generic: Matlab's FD + Q + implementation using generic
	FPGA multipliers (using the VHDL `\texttt{*}'~operator, potentially using DSP blocks);
	\item 3-step Generic NoDSP: same as above with disabled DSP blocks for synthesis and implementation;
	\item 3-step MCM: Matlab's FD + Q + optimal shift-and-add
	implementation for FxP coefficients;
	\item 2-step KCM: Matlab's FD + approach from~\cite{VolkovaIstoanDeDinechinHilaire_TowardsHardwareIIR_2019},
	which directly obtains a KCM-based implementation for real coefficients;
	\item Ours: 1-step combined approach, performing directly FD \& Q \& I using optimal shift-and-add
	multipliers.
\end{itemize}
The 3-step Generic (NoDSP) can be seen as a baseline while 3-step MCM and 2-step KCM is the state-of-the-art.
With such a setting, our goal is to analyze the benefits of the global approach compared to 2- or 3-step
approaches that first fix filter coefficients to some values and only then optimize for implementation.

\paragraph*{Bit widths} The coefficient word length is an input parameter for our tool, hence, for each
specification we will explore a range of different coefficient word lengths. For the input/output data in
hardware implementation, we used 8-, 12- and 16-bit configurations but, due to the lack of space, only
the results for 16-bit experiments are reported in the paper. See the tool's git repository for full
benchmark information.

%

\subsection{Evaluation of the ILP Model and Design Results}
In the following we evaluate the performance of our ILP model, explore the design space and compare
with the 3-step MCM-based approach to see the benefits \wrt{} number of adders in a shift-and-add
implementation.
All experiments were performed on a Linux laptop with i7-10810U processor and 32 GB RAM.
Any generic solver with an interface for \verb*|JuMP| library \cite{DunningHuchetteLubin2017} can be used, here we used CPLEX 12.10~\cite{CPLEXUsersManual_2020}.

\begin{table}[]
		\centering
		\caption{Results for our global optimization method vs. applying optimal MCM upon quantized coefficients.
			The total number of adders  $A$ ($A=A_M+A_S$) consists in the multiplier block $A_M$, and structural $A_S$ adders. 
			Results are reported for the smallest coefficient word length $W$ possible.}
	\label{tab:results_recap_hardware}
		{\small
	\begin{tabular}{@{}lllllllll@{}}
		\toprule
		\multirow{2}{*}{Benchmark}   & \multicolumn{4}{c}{Our method}             & \multicolumn{4}{c}{3-step MCM} \\ \cmidrule(l){2-5}  
		\cmidrule(lr){6-9}
		& W & $A_M$ & $A_S$ & $A$                    & W     & $A_M$  & $A_S$  & A    \\ \midrule
		\multicolumn{1}{l|}{\lpfig{1}{0}} & 4 & 1     & 4     & \multicolumn{1}{l|}{5} & 10    & 8      & 4      & 12   \\
		\multicolumn{1}{l|}{\lpfig{1}{1}} & 4 & 1     & 4     & \multicolumn{1}{l|}{5} & 16    & 11     & 4      & 15$^*$  \\
		\multicolumn{1}{l|}{\lpfig{1}{2}} & 4 & 2     & 4     & \multicolumn{1}{l|}{6} & 6     & 5      & 4      & 9    \\
		\multicolumn{1}{l|}{\lpfig{1}{3}} & 4 & 3     & 4     & \multicolumn{1}{l|}{7} & 10    & 7      & 4      & 11   \\
		\multicolumn{1}{l|}{\lpfig{1}{4}} & 5 & 4     & 4     & \multicolumn{1}{l|}{8} & 9     & 7      & 4      & 11   \\
		\multicolumn{1}{l|}{\lpfig{1}{5}} & 5 & 4     & 4     & \multicolumn{1}{l|}{8} & $-$   & $-$    & $-$    & $-$  \\
		\multicolumn{1}{l|}{\lpfig{2}{0}} & 4 & 1     & 4     & \multicolumn{1}{l|}{5} & 10    & 8      & 4      & 12   \\
		\multicolumn{1}{l|}{\lpfig{2}{1}} & 4 & 1     & 4     & \multicolumn{1}{l|}{5} & 10    & 8      & 4      & 12   \\
		\multicolumn{1}{l|}{\lpfig{2}{2}} & 5 & 3     & 4     & \multicolumn{1}{l|}{7} & 10    & 8      & 4      & 12   \\
		\multicolumn{1}{l|}{\lpfig{2}{3}} & 6 & 4     & 4     & \multicolumn{1}{l|}{8} & $-$   & $-$    & $-$    & $-$  \\
		\multicolumn{1}{l|}{\lpfig{3}{0}} & 4 & 1     & 4     & \multicolumn{1}{l|}{5} & 10    & 8      & 4      & 12   \\
		\multicolumn{1}{l|}{\lpfig{3}{1}} & 4 & 2     & 4     & \multicolumn{1}{l|}{6} & 4     & 2      & 4      & 6    \\
		\multicolumn{1}{l|}{\lpfig{3}{2}} & 4 & 3     & 4     & \multicolumn{1}{l|}{7} & 23    & 18     & 4      & 22$^*$  \\
		\multicolumn{1}{l|}{\lpfig{3}{3}} & 5 & 3     & 4     & \multicolumn{1}{l|}{7} & $-$   & $-$    & $-$    & $-$  \\
		\multicolumn{1}{l|}{lp4}     & 4 & 1     & 4     & \multicolumn{1}{l|}{5} & 4     & 3      & 4      & 7    \\
		\multicolumn{1}{l|}{hp0}     & 6 & 1     & 2     & \multicolumn{1}{l|}{3} & 11    & 6      & 4      & 10   \\
		\bottomrule		
	\end{tabular}

	\vspace{0.25\baselineskip}
{\raggedright $\phantom{.}^*$\! heuristic solution using~\cite{KummZipfFaustChang_Pipelinedaddergraph_2012} \par}
}
\end{table}

The first remark concerning our tool is that the running times are quite reasonable, varying from $10$ seconds for small word lengths ($4$-$5$~bits) and going up to a few
minutes for the largest word length we can deal with, which is roughly $10$~bits.
By default, we use the symmetry breaking constraints as in general we
observed a significant improvement (around $2\times$-$20\times$) in running times, depending on
problem
complexity. Obviously, the complexity of the ILP model is increasing with increasing the word length,
since the
ranges of integer variables are doubled with each new coefficient bit, and a few additional variables and
constraints are added as well. However, it is not the model complexity but numerical instabilities that
represent the main bottleneck in pushing the coefficient word lengths further than $10$-$11$~bits. Indeed,
our ILP model for the MCM design makes intensive use of the so-called big-M constraints, that are
limited by certain floating-point tolerances internal to the solver, beyond which the solver cannot use
efficient floating-point arithmetic for integer programming. An alternative to the big-M are the indicator
constraints but the drawback is the increased computational time, leading to a similar bound on the
maximum coefficient word lengths.

The second remark is that with our tool, an infeasibility of the design problem can be quickly proven. For
example, for the specification \lp{1}{4} in just a few seconds we prove that no implementation that
perfectly fits the specification with word
length 4 is possible. This is an important feature, since when trying to lower the coefficient word length
as much as possible, the filter designer can quickly stop the exploration. Inversely, when searching the
smallest feasible word length, the design iteration will quickly move on from infeasible ones.

The goal of our tool is to provide optimal architectures \wrt{} the number of adders in the multiplierless
implementation. The number of adders is not a fine-grained metric but it enables the design-space
exploration a priori, before any hardware synthesis and experiments. It is a good indicator of the
performance of implemented systems, as the number of adders is correlated with the number of LUTs.

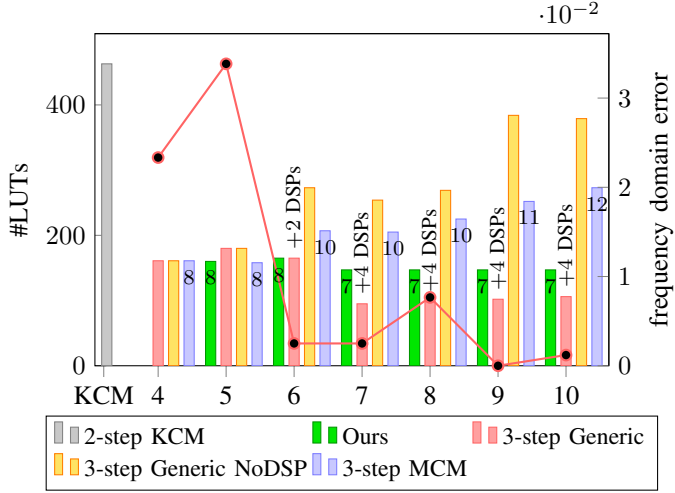
\begin{figure}
	\centering
	\def\varwidth{0.95\linewidth}
	\def\varheight{6cm}
	\begin{tikzpicture}
		\begin{axis}[
		ybar,
		bar width=0.15,
		legend style={at={(0.5,-0.15)}, anchor=north, legend columns=3, font=\small},
		legend cell align={left},
		ymin=0,
		ytick pos=left,
		xtick pos=bottom,
		width=\varwidth,
		height=\varheight,
		ylabel={\#LUTs},
		xtick={0.2, 1, 2, 3, 4, 5, 6, 7},
		xticklabels={KCM, 4, 5, 6, 7, 8, 9, 10},
		point meta=explicit symbolic
		]
		\addplot [white, only marks, draw opacity=0, fill opacity=0] coordinates {(0.7, 1) (1, 1) (2, 1) (3, 1) (4, 1) (5, 1) (6, 1) (7, 1) };
		\addplot[legend entry=2-step KCM, twostepkcm, fill=twostepkcmfill] coordinates {(0.7, 463)};
		\addplot[every node near coord/.append style={font=\footnotesize, anchor=north, color=black}, nodes near coords, nodes near coords align={vertical}, legend entry=Ours, ours, fill=oursfill] coordinates {
			(2, 160)[$8$]
			(3, 165)
			(4, 147)[$7$]
			(5, 147)[$7$]
			(6, 147)[$7$]
			(7, 147)[$7$]
		};
		\addplot[every node near coord/.append style={font=\footnotesize, rotate=90, anchor=west, color=black}, nodes near coords, nodes near coords align={vertical}, legend entry=3-step Generic, twostepgeneric, fill=twostepgenericfill] coordinates {
			(1, 161)
			(2, 180)
			(3, 165)[$+2$ DSPs]
			(4, 95)[$+4$ DSPs]
			(5, 99)[$+4$ DSPs]
			(6, 102)[$+4$ DSPs]
			(7, 106)[$+4$ DSPs]
		};
		\addplot[legend entry=3-step Generic NoDSP, twostepgenericnodsp, fill=twostepgenericnodspfill] coordinates {
			(1, 161)
			(2, 180)
			(3, 273)
			(4, 254)
			(5, 269)
			(6, 384)
			(7, 379)
		};
		\addplot[every node near coord/.append style={font=\footnotesize, anchor=north, color=black}, nodes near coords, nodes near coords align={vertical}, legend entry=3-step MCM, twostepmcm, fill=twostepmcmfill] coordinates {
			(1, 161)[$8$]
			(2, 158)[$8$]
			(3, 207)[$10$]
			(4, 205)[$10$]
			(5, 225)[$10$]
			(6, 252)[$11$]
			(7, 273)[$12$]
		};
		\end{axis}
		\begin{axis}[
		axis lines*=right,
		ymin=0.0,
		ytick pos=right,
		xmajorticks=false,
		width=\varwidth,
		height=\varheight,
		ylabel={frequency domain error},
		xtick={0.7, 1, 2, 3, 4, 5, 6, 7},
		xticklabels={KCM, 4, 5, 6, 7, 8, 9, 10},
		point meta=explicit symbolic,
		]
		\addplot[draw=white,ultra thick,draw opacity=0] coordinates {(0.7, 0) (1, 0) (2, 0) (3, 0) (4, 0) (5, 0) (6, 0) (7, 0) };
		\addplot[draw=errorplotlines,thick,mark=*,mark options={solid,draw=errorplotdots,fill=errorplotcenterdots}] coordinates {(1, 0.023333333333333) (2, 0.033835455863942) (3, 0.0025) (4, 0.0025) (5, 0.007659842074314) (6, 0.0) (7, 0.001197916666667) };
		\end{axis}
		\node at (2.45,1.21) {\footnotesize$8$};
	\end{tikzpicture}
	\caption{Implementations for \lp{1}{4} benchmark with coefficient word lengths varying from 4 to 10 bits. 
The bars correspond to the number of LUTs (left axis) and labels on bars denote number of adders. The frequency-domain error of 3-step methods is the red line (right axis).}
	\label{fig:lp1x4dw16_luts}
\end{figure}

We claim that optimizing the filter coefficients simultaneously for the filter design and MCM problem
yields smaller total number of adders than the 3-step MCM approach, as it covers the whole design
space.
Table~\ref{tab:results_recap_hardware} presents the optimization results for our method, and for the
3-step MCM method. 
We report results for the smallest word lengths possible, in which our method has a feasible result (and, by construction, no frequency domain error) and the Q step in 3-step MCM yields stable filters with frequency-domain error smaller
than~$10^{-7}$.

It can be noticed that our method finds solutions for smaller coefficient word lengths than the 3-step
MCM, and with significant smaller total number of adders for most of the cases ($\approx\!\! 46
\%$ on average in Table~\ref{tab:results_recap_hardware}). In some cases, marked
with asterisk, target word lengths for 3-step MCM were too big \st{} the optimal MCM timed-out and the
RPAG heuristic~\cite{KummZipfFaustChang_Pipelinedaddergraph_2012} was used to obtain adder graphs instead.
For the specification \filtername{lp4} both approaches find designs with 4-bit coefficients, but thanks to
efficiently
covering the whole design space of all possible FxP coefficients, our method determines
coefficients that require fewer adders in the multiplier block and the total of $5$ adders, instead of $7$.
Finally, the benchmark \lp{3}{1} is the only case when the coefficients in 3-step MCM coincide
with the ones found by our tool.

Another advantage of the proposed ILP formulation is privileging sparse implementations, which even for
second-order filters largely improves performance due to smaller number of structural adders. For
instance, the benchmark \filtername{hp0} was traditionally implemented with all non-zero
coefficients~\cite{VolkovaIstoanDeDinechinHilaire_TowardsHardwareIIR_2019,
SarbisheiRadeckaZilic_AnalyticalOptimizationBit_2012}. We have approximated the frequency response
of this compensator using the poles and zeros from the original
paper~\cite{KrachFrackeltonCarlettaVeillette_FPGAbasedimplementation_2003} and used it as reference
in our ILP. As a result, a sparse implementation with 6-bit coefficients was possible, having as
coefficients $ b_0 = 1$, $b_1 = -1$, $b_2 = 0$, $a_1 = -31/32$, $a_2 = 0$ and leading to the
total of $3$ adders for the whole filter. To compare, the 3-step MCM can provide at minimum 11-bit coefficients implemented with $10$ adders.
Moreover, our design has poles further apart from the unit circle, which improves the numerical quality of the
time-domain implementation.

Our tool provides results for difficult specifications even when FD with Matlab fails. For benchmarks
\lp{1}{5}, \lp{2}{3} and
\lp{3}{3} no IIR filter design method in Matlab could find double-precision coefficients for a frequency
response of sufficient quality. Our method, however, successfully completes the task.




It is important to note that our tool provides optimal implementations \wrt{} the total number of adders
\textit{for a given coefficient word length.} In other words, finding the best coefficient word length is still
the filter designer task. While typical flow is to stop at the smallest possible coefficient word length, as in
Table~\ref{tab:results_recap_hardware}, it
does not necessarily lead to the optimal implementation \wrt{} all possible word lengths, and increasing
the coefficient word length can actually lead to fewer adders.

To illustrate this, \figurename~\ref{fig:lp1x4dw16_luts} shows synthesis results for the \lp{1}{4} benchmark with
coefficient
word lengths varying from 4 to 10 bits (the hardware implementation is described in details in
Section~\ref{sec:hardware}).
It can be seen that increasing the word length from 5 bits to 7 bits actually reduces
the total number of adders from 8 to 7 and improves the result implementation. 
Moreover, our approach provides a stable behavior: once a 7-adder solution is found, increasing coefficient word length
will never lead to a worse implementation. In practice, the ILP either finds different coefficients with the same number of adders, or simply multiplies values by two for each additional bit (which is the case in \figurename~\ref{fig:lp1x4dw16_luts} starting 7 bits). In other words, even if the user specifies a larger word length than required, our tool finds the best coefficients
that might fit in a smaller format and guarantees that trying smaller word length will not give smaller number of adders.

For the classical 3-step approaches, design-space exploration is more difficult and irregular.
For these methods, coefficient quantization introduces a frequency-domain error meaning that the quantized filter does not satisfy the frequency specifications any more (see the red line in \figurename~\ref{fig:lp1x4dw16_luts}). This error is highly non-linear, and a typical intuition that increasing coefficient word length improves the quality of filter is simply not true (see the frequency-domain error for 7- and 8-bit coefficients in \figurename~\ref{fig:lp1x4dw16_luts}). 
Hence, the search for the best coefficient size must be exhaustive for 3-step methods.

\subsection{Hardware Implementation and Discussion}\label{sec:hardware}


In the following, we first describe in details the  faithfully-rounded architectures that we implement in FloPoCo and then
discuss the synthesis results obtained for our benchmarks and each approach.

\paragraph*{Implemented architecture}

We aim at providing faithfully-rounded implementations, \ie{}, the precision of the output $y$ (the Least Significant Bit (LSB) position $\ell_\text{out}$) serves as the accuracy constraint.
There are different ways to assign the sizes of data paths for a filter implementation, and it is important to
not underestimate the sizes (otherwise overflow occurs) but, on the other hand, assigning larger data
sizes will waste resources to compute unnecessary bits.
Hence, our goal is to provide a code generator that, given the input/output format and filter coefficients,
automatically computes the word sizes of all internal data paths to guarantee the time-domain error
smaller than $2^{\ell_\text{out}}$ but not more.

\begin{figure}
	\centering
	\begin{tikzpicture}
	\matrix (secondorderiir) [row sep=2mm, column sep=-1.7mm] {
		\node[coordinate] (m1x0) {}; \pgfmatrixnextcell
		\node[dspnodeopen,dsp/label=left] (m1x1) {$\underset{(m_{\inm}, \ell_{\inm})}{x}$\!\!\!}; \pgfmatrixnextcell
		\node[coordinate] (m1x3) {}; \pgfmatrixnextcell
		\node[coordinate] (m1x4) {}; \pgfmatrixnextcell
		\node[coordinate, label={left:$\phantom{.}$}] (m1x5) {}; \pgfmatrixnextcell
		\node[coordinate] (m1x6) {}; \pgfmatrixnextcell
		\node[dspadder, label={right:$\phantom{.}$}] (m1x7) {}; \pgfmatrixnextcell
		\node[dspsquare, scale=0.7] (m1x8) {\large $\,<\!\!< \ell_b\,$}; \pgfmatrixnextcell
		\node[dspadder, label={left:$\phantom{..}$}, label={right:$\phantom{.}$}] (m1x9) {}; \pgfmatrixnextcell
		\node[dspsquare, scale=0.7] (m1x10) {\large $\,<\!\!< \ell_a\,$}; \pgfmatrixnextcell
		\node[coordinate, label={[yshift=0.25cm]above:$\phantom{..}\ell_{\extm}$}] (m1x11) {}; \pgfmatrixnextcell
		\node[dspnodefull] (m1x12) {}; \pgfmatrixnextcell
		\node[dspnodeopen, label={[yshift=0.2cm]above:$\ell_{\outm}\phantom{.....}$},dsp/label=rigth, label=left:$\phantom{.}$] (m1x13) {$\underset{(m_{\outm}, \ell_{\outm})}{y}$}; \\
		\node[coordinate] (m2x1) {}; \pgfmatrixnextcell
		\node[coordinate, label={below:$\phantom{2}\leftarrow 2$}] (m2x2) {}; \pgfmatrixnextcell
		\node[dspnodefull] (m2x3) {}; \pgfmatrixnextcell
		\node[coordinate] (m2x4) {}; \pgfmatrixnextcell
		\node[coordinate] (m2x5) {}; \pgfmatrixnextcell
		\node[coordinate] (m2x6) {}; \pgfmatrixnextcell
		\node[coordinate] (m2x7) {}; \pgfmatrixnextcell
		\node[coordinate] (m2x8) {}; \pgfmatrixnextcell
		\node[coordinate] (m2x9) {}; \pgfmatrixnextcell
		\node[coordinate] (m2x10) {}; \pgfmatrixnextcell
		\node[coordinate] (m2x11) {}; \pgfmatrixnextcell
		\node[coordinate] (m2x12) {}; \pgfmatrixnextcell
		\node[coordinate] (m2x13) {}; \\
		\node[coordinate] (m3x1) {}; \pgfmatrixnextcell
		\node[coordinate] (m3x2) {}; \pgfmatrixnextcell
		\node[dspadder, label={below right:$5x$}] (m3x3) {}; \pgfmatrixnextcell
		\node[coordinate] (m3x4) {}; \pgfmatrixnextcell
		\node[coordinate] (m3x5) {}; \pgfmatrixnextcell
		\node[coordinate] (m3x6) {}; \pgfmatrixnextcell
		\node[dspsquare, scale=0.7] (m3x7) {$z^{-1}$}; \pgfmatrixnextcell
		\node[coordinate] (m3x8) {}; \pgfmatrixnextcell
		\node[dspsquare, scale=0.7] (m3x9) {$z^{-1}$}; \pgfmatrixnextcell
		\node[coordinate, label={below:$\phantom{2}\leftarrow 2$}] (m3x10) {}; \pgfmatrixnextcell
		\node[dspnodefull] (m3x11) {}; \pgfmatrixnextcell
		\node[coordinate] (m3x12) {}; \pgfmatrixnextcell
		\node[coordinate] (m3x13) {}; \\
		\node[coordinate] (m4x1) {}; \pgfmatrixnextcell
		\node[coordinate] (m4x2) {}; \pgfmatrixnextcell
		\node[dspnodefull] (m4x3) {}; \pgfmatrixnextcell
		\node[coordinate] (m4x4) {}; \pgfmatrixnextcell
		\node[coordinate] (m4x5) {}; \pgfmatrixnextcell
		\node[coordinate, label={[yshift=-0.7mm]below:$\leftarrow$}, label={below right:$\ 3$}] (m4x6) {}; \pgfmatrixnextcell
		\node[coordinate] (m4x7) {}; \pgfmatrixnextcell
		\node[coordinate] (m4x8) {}; \pgfmatrixnextcell
		\node[coordinate] (m4x9) {}; \pgfmatrixnextcell
		\node[coordinate] (m4x10) {}; \pgfmatrixnextcell
		\node[dspadder, label={below right:$5y$}] (m4x11) {}; \pgfmatrixnextcell
		\node[coordinate] (m4x12) {}; \pgfmatrixnextcell
		\node[coordinate] (m4x13) {}; \\
		\node[coordinate] (m5x1) {}; \pgfmatrixnextcell
		\node[coordinate, label={below:$\phantom{2}\leftarrow 2$}] (m5x2) {}; \pgfmatrixnextcell
		\node[dspnodefull] (m5x3) {}; \pgfmatrixnextcell
		\node[coordinate] (m5x4) {}; \pgfmatrixnextcell
		\node[coordinate] (m5x5) {}; \pgfmatrixnextcell
		\node[coordinate] (m5x6) {}; \pgfmatrixnextcell
		\node[dspadder] (m5x7) {}; \pgfmatrixnextcell
		\node[coordinate] (m5x8) {}; \pgfmatrixnextcell
		\node[dspadder, dsp/label=above right] (m5x9) {$\phantom{..}\overline{\phantom{l}}$}; \pgfmatrixnextcell
		\node[coordinate, label={below:$\phantom{3}\leftarrow 3$}] (m5x10) {}; \pgfmatrixnextcell
		\node[coordinate] (m5x11) {}; \pgfmatrixnextcell
		\node[coordinate] (m5x12) {}; \pgfmatrixnextcell
		\node[coordinate] (m5x13) {}; \\
		\node[coordinate] (m6x1) {}; \pgfmatrixnextcell
		\node[coordinate] (m6x2) {}; \pgfmatrixnextcell
		\node[dspadder, label={below right:$25x$}] (m6x3) {}; \pgfmatrixnextcell
		\node[coordinate] (m6x4) {}; \pgfmatrixnextcell
		\node[coordinate] (m6x5) {}; \pgfmatrixnextcell
		\node[coordinate] (m6x6) {}; \pgfmatrixnextcell
		\node[dspsquare, scale=0.7] (m6x7) {$z^{-1}$}; \pgfmatrixnextcell
		\node[coordinate] (m6x8) {}; \pgfmatrixnextcell
		\node[dspsquare, scale=0.7] (m6x9) {$z^{-1}$}; \pgfmatrixnextcell
		\node[dspnodefull] (m6x10) {}; \pgfmatrixnextcell
		\node[coordinate] (m6x11) {}; \pgfmatrixnextcell
		\node[coordinate] (m6x12) {}; \pgfmatrixnextcell
		\node[coordinate] (m6x13) {}; \\
		\node[coordinate] (m7x1) {}; \pgfmatrixnextcell
		\node[coordinate] (m7x2) {}; \pgfmatrixnextcell
		\node[coordinate] (m7x3) {}; \pgfmatrixnextcell
		\node[coordinate] (m7x4) {}; \pgfmatrixnextcell
		\node[dspnodefull, label=above:$\phantom{...}$] (m7x5) {}; \pgfmatrixnextcell
		\node[coordinate] (m7x6) {}; \pgfmatrixnextcell
		\node[coordinate] (m7x7) {}; \pgfmatrixnextcell
		\node[coordinate] (m7x8) {}; \pgfmatrixnextcell
		\node[coordinate] (m7x9) {}; \pgfmatrixnextcell
		\node[coordinate, label={above:$\phantom{2}\leftarrow 2$}] (m7x10) {}; \pgfmatrixnextcell
		\node[coordinate] (m7x11) {}; \pgfmatrixnextcell
		\node[coordinate] (m7x12) {}; \pgfmatrixnextcell
		\node[coordinate] (m7x13) {}; \\
	};
	
	\begin{scope}[start chain]
	\chainin (m1x1);
	\chainin (m1x3) [join=by dspline];
	\chainin (m2x3) [join=by dspline];
	\chainin (m2x2) [join=by dspline];
	\chainin (m3x2) [join=by dspline];
	\chainin (m3x3) [join=by dspconn];
	\end{scope}
	\begin{scope}[start chain]
	\chainin (m2x3);
	\chainin (m2x4) [join=by dspline];
	\chainin (m3x4) [join=by dspline];
	\chainin (m3x3) [join=by dspconn];
	\end{scope}
	\begin{scope}[start chain]
	\chainin (m3x3);
	\chainin (m4x3) [join=by dspline];
	\chainin (m4x6) [join=by dspline];
	\chainin (m5x6) [join=by dspline];
	\chainin (m5x7) [join=by dspconn];
	\end{scope}
	\begin{scope}[start chain]
	\chainin (m4x3);
	\chainin (m5x3) [join=by dspline];
	\chainin (m5x2) [join=by dspline];
	\chainin (m6x2) [join=by dspline];
	\chainin (m6x3) [join=by dspconn];
	\end{scope}
	\begin{scope}[start chain]
	\chainin (m5x3);
	\chainin (m5x4) [join=by dspline];
	\chainin (m6x4) [join=by dspline];
	\chainin (m6x3) [join=by dspconn];
	\end{scope}
	\begin{scope}[start chain]
	\chainin (m6x3);
	\chainin (m7x3) [join=by dspline];
	\chainin (m7x5) [join=by dspline];
	\chainin (m1x5) [join=by dspline];
	\chainin (m1x7) [join=by dspconn];
	\end{scope}
	\begin{scope}[start chain]
	\chainin (m7x5);
	\chainin (m7x7) [join=by dspline];
	\end{scope}
	\begin{scope}[start chain]
	\chainin (m1x7);
	\chainin (m1x8) [join=by dspline];
	\end{scope}
	\begin{scope}[start chain]
	\chainin (m1x8);
	\chainin (m1x9) [join=by dspconn];
	\end{scope}
	\begin{scope}[start chain]
	\chainin (m1x9);
	\chainin (m1x10) [join=by dspline];
	\end{scope}
	\begin{scope}[start chain]
	\chainin (m1x10);
	\chainin (m1x12) [join=by dspline];
	\chainin (m1x13) [join=by dspconn];
	\end{scope}
	\begin{scope}[start chain]
	\chainin (m1x10);
	\chainin (m1x12) [join=by Truncature];
	\end{scope}
	\begin{scope}[start chain]
	\chainin (m1x12);
	\chainin (m1x13) [join=by Truncature];
	\end{scope}
	\begin{scope}[start chain]
	\chainin (m1x12);
	\chainin (m2x12) [join=by dspline];
	\chainin (m2x11) [join=by dspline];
	\chainin (m3x11) [join=by dspline];
	\chainin (m3x10) [join=by dspline];
	\chainin (m4x10) [join=by dspline];
	\chainin (m4x11) [join=by dspconn];
	\end{scope}
	\begin{scope}[start chain]
	\chainin (m3x11);
	\chainin (m3x12) [join=by dspline];
	\chainin (m4x12) [join=by dspline];
	\chainin (m4x11) [join=by dspconn];
	\end{scope}
	\begin{scope}[start chain]
	\chainin (m4x11);
	\chainin (m6x11) [join=by dspline];
	\chainin (m6x10) [join=by dspline];
	\chainin (m5x10) [join=by dspline];
	\chainin (m5x9) [join=by dspconn];
	\end{scope}
	\begin{scope}[start chain]
	\chainin (m6x10);
	\chainin (m7x10) [join=by dspline];
	\chainin (m7x9) [join=by dspline];
	\end{scope}
	
	\begin{scope}[start chain]
	\chainin (m7x7);
	\chainin (m6x7) [join=by dspline];
	\end{scope}
	\begin{scope}[start chain]
	\chainin (m6x7);
	\chainin (m5x7) [join=by dspconn];
	\end{scope}
	\begin{scope}[start chain]
	\chainin (m5x7);
	\chainin (m3x7) [join=by dspline];
	\end{scope}
	\begin{scope}[start chain]
	\chainin (m3x7);
	\chainin (m1x7) [join=by dspconn];
	\end{scope}
	\begin{scope}[start chain]
	\chainin (m7x9);
	\chainin (m6x9) [join=by dspline];
	\end{scope}
	\begin{scope}[start chain]
	\chainin (m6x9);
	\chainin (m5x9) [join=by dspconn];
	\end{scope}
	\begin{scope}[start chain]
	\chainin (m5x9);
	\chainin (m3x9) [join=by dspline];
	\end{scope}
	\begin{scope}[start chain]
	\chainin (m3x9);
	\chainin (m1x9) [join=by dspconn];
	\end{scope}
	\end{tikzpicture}
	\caption{The \lp{1}{4} benchmark can be implemented with mere $7$ adders. All additions are exact, the truncation to internal
		extended
		format $\ell_{\extm}$ is determined using the WCPG
		s.t. the output is faithfully rounded to $\ell_{\outm}$.
	}
	\label{fig:lp1x4hw}
\end{figure}
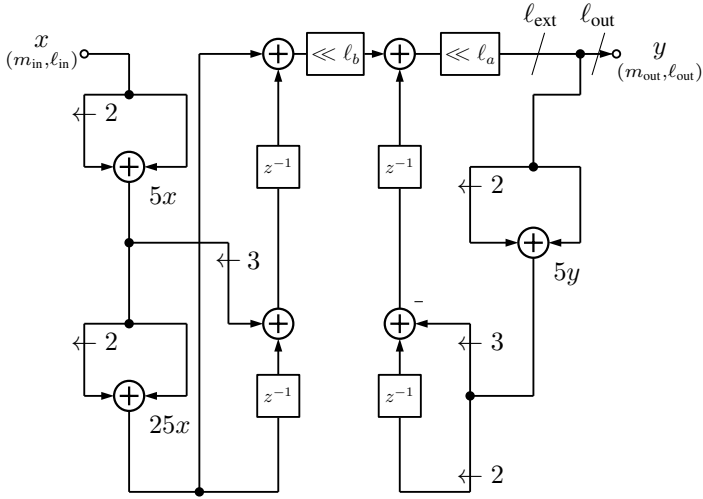

\figurename~\ref{fig:lp1x4hw} presents our approach for multiplierless hardware IIR on the example of
\lp{1}{4}
benchmark, which can be implemented with $7$ adders.
Its transfer function was obtained~as
\begin{equation}
H_{\texttt{lp}1_4}\!\left(z\right) = \frac{{25}\times{2^{-7}} + {40}\times{2^{-7}}\, z^{-1} + {25}\times{2^{-7}}\, z^{-2}}{1 - {40}\times{2^{-6}}\, z^{-1} + {20}\times{2^{-6}}\, z^{-2}}.
\end{equation}
The inputs to the architecture generator are the MSB and LSB positions of the input $x$ and output $y$,
the adder graphs for multiplier blocks $a_k$ and $b_k$, and their corresponding LSBs.
For instance, here $\ell_b = -7$, $\ell_a = -6$.

Obviously, one cannot compute exactly (or with some fixed precision) on each iteration, truncate to
$\ell_\text{out}$ and simply feed
truncated values back into the loop, as this will degrade tremendously numerical quality and
accumulated errors will explode.
The Worst-Case Peak Gain (WCPG) measure for IIR
filters~\cite{VolkovaHilaireLauter_ReliableEvaluationWorst_2015}, which has been applied for hardware
IIR filters implemented with KCM
multipliers~\cite{VolkovaIstoanDeDinechinHilaire_TowardsHardwareIIR_2019}, permits to determine the
necessary extended internal precision $\ell_\text{ext}$ s.t. the propagated error never reaches the LSB of
the output. For example, in \figurename~\ref{fig:lp1x4hw}
$\ell_{\extm} = \ell_{\outm}+G$,
where the number of guard bits $G$ for the filter \lp{1}{4} determined with its WCPG is $G=3$.
Then, the output of the multiplier blocks needs to guarantee its result with accuracy $\ell_{\extm}$. In our architecture we perform all additions and shifts exactly, increasing the size of data paths until their truncation to $\ell_{\extm}$. 


For the generic approach, based on plain VHDL multipliers (using the \texttt{*} operator), we adopt a similar approach.

%


\paragraph*{Synthesis results}

\input{images/recap.tex}

Synthesis was performed using Vivado v2019.1 for a Kintex~7 device \mbox{(xc7k70tfbv484-3)}.
The delay is reported after place and route.
We performed the experiments for 8-, 12- and 16-bit inputs/outputs considering the inputs to have the MSB position $-1$. 
For conciseness, we report the results only for the 16-bit experiments but the same observations hold for 8- and 12-bit ones.


To leave the process of (non-optimal) search for the best coefficient size out of scope, we further { compare the designs with the smallest coefficient word lengths, for which our ILP has a feasible solution, and the 3-step approach has a frequency-domain error smaller than $10^{-7}$. }
\figurename~\ref{fig:recap_all_filters} summarizes the obtained resources (LUTs and DSPs) and the critical path delay. For each benchmark, on x-axis we see the coefficient word length for our result (left) and for the 3-step quantization (right). For instance, we recognize the values 5 and 9 for the~\lp{1}{4}.

We can observe that our method is always superior to any of the classical or state-of-the-art methods, both in terms of LUTs and delay. On our benchmarks we observed the average LUT improvement of $42\%$ compared to the best results of other methods. In addition to that, the proposed approach offers the lowest delay with an
improvement of $21\%$, which is not as drastic as LUT improvement but expected due to the similar number of  {delays}. Of course, our method is more efficient partly due to the smallest possible coefficient word lengths, but even when using the same word sizes (see benchmark \lp{4}{}) our design requires less LUTs and has smaller delay.

The KCM-based approach, representing the state-of-the-art approach for faithfully-rounded IIR filters, had worse general performance than our method in all benchmarks. While the LUT consumption of KCM-based IIR filters is significantly bigger than for our approach, it is comparable to the 3-step Generic NoDSP approach. However, in terms of delay, KCM-based multipliers prove to be generally faster or comparable to 3-step methods. It should be noted that the superiority of our approach over KCM-based IIR filters is not a surprise for the small word lengths, according to the recent result~\cite{dffk19}, and while it can be expected that our approach is possible only for word lengths less than 16 bits due to optimization time-out, the KCM-based multipliers will have no issue dealing with large word lengths.

It can be noted that with the increase in frequency specification difficulty, the 3-step methods generally degrade in performance, while our design methods have a more regular behavior, providing small and fast implementations even for complicated filters.

One of the best performance improvements was achieved for the \filtername{hp0} compensator. This is due to the much smaller coefficient word length than in all the previous literature, the sparsity and the fact that we succeeded in finding a filter with poles further from the unit circle. As a consequence, the number of internal guard bits was smaller too, resulting in only $105$ LUTs compared to $286$ in the 3-step MCM and $760$ in the KCM-based filters.

\section{Conclusion and Perspectives}

We proposed a new method called \filtername{IIRoptim} for the optimal design of multiplierless second-order IIR filters w.r.t. the number of adders. Our approach is based on a combined global optimization problem, which searches for stable filter coefficients in FxP format such that the number of adders in a shift-and-add implementation is minimized.
Furthermore, we proposed an automated tool which combines \filtername{IIRoptim} with FloPoCo and provides automatic hardware code generation for implementation on FPGA. With our tool, the instabilities and quantization effects on second-order IIR filters become a thing of the past.

We proposed a linearized formulation of the combined filter design and MCM problem as one ILP model, which provides a convenient way for extensions. Several design space reduction techniques were proposed, including a novel symmetry breaking constraint, which we formally proved. As a result, the filter design and optimization takes mere seconds.

Our approach has several useful and important consequences. First, with our ILP it is easy and quick to prove the impossibility of a design, with a given coefficient word length and large enough range of MSBs, such that the filter specifications are \emph{fully} satisfied. Moreover, this would be true for any smaller coefficient word length, providing a filter designer the unprecedented assurance that the design space exploration in that direction can be stopped.
Second, if the filter design is possible with a certain given word length, increasing the word length will never yield a larger number of adders, and often the coefficients will remain the same or shifted by one bit.
With this property and the fact that all our designs are guaranteed to have zero frequency-domain error, typical non-linearities in the second-order IIR design process are no longer an issue, making the design-space exploration a more regular process.

For hardware experiments, we provided a faithfully-rounded multiplierless operator for IIR filters within FloPoCo.
The synthesis results confirmed that a global optimization approach is superior to the multi-step FD \& Q \& I classical methods, and even to the state-of-the-art KCM-based IIR filters. With the ILP formulation we search directly the FxP coefficient design space, privileging sparse implementations and sometimes finding the filters that even double-precision Matlab fails to find. After testing the tool on numerous benchmarks, we observed a 42\% improvement in number of LUTs, and 21\% improvement in delay, on average.

The superiority over the KCM-based IIR~\cite{VolkovaIstoanDeDinechinHilaire_TowardsHardwareIIR_2019}, that were first to introduce faithfully-rounded filters using analysis of the worst-case rounding errors, advances the progress towards reliable IIR filters and demonstrates again that numerical guarantees do not necessarily come at a higher cost.

Some efforts are still required to extend our method, in particular to higher order filters. We see two possible
directions for that extension, first, a single ILP model which would permit the design of cascaded
second order sections and, second, an external loop for the decomposition of specifications into simpler
specifications that are reachable by a second order filter. Extension to other structures than the Direct Forms is also a promising direction that should be tackled in the future.

Although the number of adders is a reliable high level metric, optimizing the number full adders instead
would be better. We are fairly optimistic on the fact that our method can be refined to minimize that
criteria, first as a post-design optimization of truncations, and then as one global optimization problem.
{Furthermore, we plan to introduce the truncations of data paths and model the rounding-error in the ILP model.
This will permit further performance gains in the implemented filters.}


%
%
%
%


\bibliographystyle{IEEEtran}
\bibliography{iiroptim}

\begin{thebibliography}{10}
\providecommand{\url}[1]{#1}
\csname url@samestyle\endcsname
\providecommand{\newblock}{\relax}
\providecommand{\bibinfo}[2]{#2}
\providecommand{\BIBentrySTDinterwordspacing}{\spaceskip=0pt\relax}
\providecommand{\BIBentryALTinterwordstretchfactor}{4}
\providecommand{\BIBentryALTinterwordspacing}{\spaceskip=\fontdimen2\font plus
\BIBentryALTinterwordstretchfactor\fontdimen3\font minus
  \fontdimen4\font\relax}
\providecommand{\BIBforeignlanguage}[2]{{%
\expandafter\ifx\csname l@#1\endcsname\relax
\typeout{** WARNING: IEEEtran.bst: No hyphenation pattern has been}%
\typeout{** loaded for the language `#1'. Using the pattern for}%
\typeout{** the default language instead.}%
\else
\language=\csname l@#1\endcsname
\fi
#2}}
\providecommand{\BIBdecl}{\relax}
\BIBdecl

\bibitem{SmekalVich_OptimizedmodelsIIR_1999}
Z.~Sm\'{e}kal and R.~V\'{i}ch, ``{Optimized models of IIR digital filters for
  fixed-point digital signal processor},'' in \emph{ICECS'99. Proceedings of
  ICECS '99. 6th IEEE International Conference on Electronics, Circuits and
  Systems}, vol.~1, Sep. 1999, pp. 145--148.

\bibitem{VanuytselBoetsVanBiesenTemmerman_Efficienthybridoptimization_2002}
G.~Vanuytsel, P.~Boets, L.~Van~Biesen, and S.~Temmerman, ``{Efficient hybrid
  optimization of fixed-point cascaded IIR filter coefficients},'' in
  \emph{IMTC/2002. Proceedings of the 19th IEEE Instrumentation and Measurement
  Technology Conference}, vol.~1, May 2002, pp. 793--797.

\bibitem{WangLiLi_Fixedpointdigital_2010}
Y.~Wang, B.~Li, and Z.~Li, ``{Fixed-point digital IIR filter design using
  multi-objective optimization evolutionary algorithm},'' in \emph{2010 IEEE
  Youth Conference on Information, Computing and Telecommunications}, Nov.
  2010, pp. 174--177.

\bibitem{GeversLi93}
M.~Gevers and G.~Li, \emph{Parametrizations in Control, Estimation and
  Filtering Problems: Accuracy Aspects}, 01 1993.

\bibitem{dm95a}
A.~Dempster and M.~Macleod, ``{Use of Minimum-Adder Multiplier Blocks in FIR
  Digital Filters},'' \emph{IEEE Transactions on Circuits and Systems II:
  Analog and Digital Signal Processing}, vol.~42, no.~9, pp. 569--577, 1995.

\bibitem{vp07}
Y.~Voronenko and M.~P{\"u}schel, ``{Multiplierless Multiple Constant
  Multiplication},'' \emph{ACM Transactions on Algorithms}, vol.~3, no.~2, pp.
  1--38, 2007.

\bibitem{KummZipfFaustChang_Pipelinedaddergraph_2012}
M.~Kumm, P.~Zipf, M.~Faust, and C.-H. Chang, ``{Pipelined adder graph
  optimization for high speed multiple constant multiplication},'' in
  \emph{2012 {IEEE} International Symposium on Circuits and Systems}, May 2012.

\bibitem{acfm08}
L.~Aksoy, E.~da~Costa, P.~Flores, and J.~Monteiro, ``{Exact and Approximate
  Algorithms for the Optimization of Area and Delay in Multiple Constant
  Multiplications},'' \emph{IEEE Transactions on Computer-Aided Design of
  Integrated Circuits and Systems}, vol.~27, no.~6, pp. 1013--1026, 2008.

\bibitem{kfmzm13}
M.~Kumm, D.~Fangh{\"a}nel, K.~M{\"o}ller, P.~Zipf, and U.~Meyer-Baese, ``{FIR
  Filter Optimization for Video Processing on FPGAs},'' \emph{Springer EURASIP
  Journal on Advances in Signal Processing}, pp. 1--18, 2013.

\bibitem{Kumm_MultipleConstantMultiplication_2016_book}
M.~Kumm, \emph{{Multiple Constant Multiplication Optimizations for Field
  Programmable Gate Arrays}}.\hskip 1em plus 0.5em minus 0.4em\relax Springer
  Fachmedien Wiesbaden, 2016.

\bibitem{Kumm_OptimalConstantMultiplication_2018}
------, ``{Optimal Constant Multiplication Using Integer Linear Programming},''
  \emph{IEEE Transactions on Circuits and Systems II: Express Briefs}, vol.~65,
  no.~5, pp. 567--571, 2018.

\bibitem{c94}
K.~D. Chapman, ``{Fast Integer Multipliers Fit in FPGAs},'' \emph{Electronic
  Design News}, 1994.

\bibitem{fc11}
M.~Faust and C.-H. Chang, ``{Bit-parallel Multiple Constant Multiplication
  using Look-Up Tables on FPGA},'' \emph{IEEE International Symposium of
  Circuits and Systems (ISCAS)}, pp. 657--660, 2011.

\bibitem{dffk19}
F.~d. Dinechin, S.-I. Filip, L.~Forget, and M.~Kumm, ``{Table-Based versus
  Shift-And-Add Constant Multipliers for FPGAs},'' in \emph{IEEE Symposium on
  Computer Arithmetic (ARITH)}, 2019.

\bibitem{VolkovaIstoanDeDinechinHilaire_TowardsHardwareIIR_2019}
A.~Volkova, M.~Istoan, F.~De~Dinechin, and T.~Hilaire, ``{Towards Hardware IIR
  Filters Computing Just Right: Direct Form I Case Study},'' \emph{IEEE
  Transactions on Computers}, vol.~68, no.~4, pp. 597--608, 2019.

\bibitem{Opp2009}
A.~V. Oppenheim and R.~W. Schafer, \emph{Discrete-Time Signal Processing},
  3rd~ed.\hskip 1em plus 0.5em minus 0.4em\relax USA: Prentice Hall Press,
  2009.

\bibitem{Leich_ToolboxdesignIIR_1997}
H.~Leich, ``{Toolbox for the design of IIR digital filters},'' in
  \emph{Proceedings of 13th International Conference on Digital Signal
  Processing}, vol.~2, Jul. 1997, pp. 621--624.

\bibitem{JenkinsNayeri_Adaptivefiltersrealized_1986}
W.~K. Jenkins and M.~Nayeri, ``Adaptive filters realized with second order
  sections,'' in \emph{{ICASSP} '86. {IEEE} International Conference on
  Acoustics, Speech, and Signal Processing}.\hskip 1em plus 0.5em minus
  0.4em\relax Institute of Electrical and Electronics Engineers, 1986.

\bibitem{DempsterMacleod_ComparisonIIRfilter_1995}
A.~G. Dempster and M.~D. Macleod, ``{Comparison of IIR filter structure
  complexities using multiplier blocks},'' in \emph{Proceedings of ISCAS'95 -
  International Symposium on Circuits and Systems}, vol.~2, Apr. 1995, pp.
  858--861.

\bibitem{CuckooSearch}
S.~Ansari, G.~Kishor, P.~K. Verma, N.~Agrawal, I.~Sharma, and A.~Kumar,
  ``Design of multiplierless digital iir filter using modified cuckoo search
  algorithm,'' in \emph{2018 International Conference on Communication and
  Signal Processing (ICCSP)}, 2018, pp. 0405--0410.

\bibitem{DesignOptimalMultiplierless_2019}
\BIBentryALTinterwordspacing
M.~Kumm, A.~Volkova, and S.-I. Filip, ``{Design of Optimal Multiplierless FIR
  Filters with Minimal Number of Adders},'' May 2021, working paper or
  preprint. [Online]. Available:
  \url{https://hal.archives-ouvertes.fr/hal-02392522}
\BIBentrySTDinterwordspacing

\bibitem{Hettich1993}
R.~Hettich and K.~O. Kortanek, ``Semi-infinite programming: Theory, methods,
  and applications,'' \emph{SIAM Review}, vol.~35, no.~3, pp. 380--429, 1993.

\bibitem{VolkovaLauterHilaire_ReliableVerificationDigital_2017}
A.~Volkova, C.~Lauter, and T.~Hilaire, ``{Reliable Verification of Digital
  Implemented Filters Against Frequency Specifications},'' in \emph{2017 {IEEE}
  24th Symposium on Computer Arithmetic ({ARITH})}.\hskip 1em plus 0.5em minus
  0.4em\relax {IEEE}, Jul. 2017.

\bibitem{Marendet2020}
A.~Marendet, A.~Goldsztejn, G.~Chabert, and C.~Jermann, ``A standard
  branch-and-bound approach for nonlinear semi-infinite problems,''
  \emph{EJOR}, vol. 282, no.~2, pp. 438--452, 2020.

\bibitem{BillionnetElloumiLambert_LinearReformulationsInteger_2008}
A.~Billionnet, S.~Elloumi, and A.~Lambert, ``{Linear Reformulations of Integer
  Quadratic Programs},'' in \emph{Modelling, Computation and Optimization in
  Information Systems and Management Sciences}.\hskip 1em plus 0.5em minus
  0.4em\relax Berlin, Heidelberg: Springer Berlin Heidelberg, 2008, pp. 43--51.

\bibitem{Glover_ImprovedLinearInteger_1975}
F.~Glover, ``{Improved Linear Integer Programming Formulations of Nonlinear
  Integer Problems},'' \emph{Management Science}, vol.~22, no.~4, pp. 455--460,
  Dec. 1975.

\bibitem{OralKettani_LinearizationProcedureQuadratic_1992}
M.~Oral and O.~Kettani, ``{A Linearization Procedure for Quadratic and Cubic
  Mixed-Integer Problems},'' \emph{Operations Research}, vol.~40, no.
  1-supplement-1, pp. S109--S116, Feb. 1992.

\bibitem{BertsimasTsitsiklis_Introductionlinearoptimization_1997_book}
D.~Bertsimas and J.~Tsitsiklis, \emph{Introduction to linear
  optimization}.\hskip 1em plus 0.5em minus 0.4em\relax Belmont, Mass: Athena
  Scientific, 1997.

\bibitem{Mangasarian_Absolutevalueprogramming_2006}
O.~L. Mangasarian, ``Absolute value programming,'' \emph{Computational
  Optimization and Applications}, vol.~36, no.~1, pp. 43--53, Nov. 2006.

\bibitem{Antoniou_DigitalFiltersAnalysis_2018_book}
A.~Antoniou, \emph{{Digital Filters: Analysis, Design, and Signal Processing
  Applications}}.\hskip 1em plus 0.5em minus 0.4em\relax New York: McGraw-Hill
  Education, 2018.

\bibitem{Walsh_GeneralSymmetryBreaking_2006}
T.~Walsh, ``{General Symmetry Breaking Constraints},'' in \emph{Principles and
  Practice of Constraint Programming - {CP} 2006}.\hskip 1em plus 0.5em minus
  0.4em\relax Springer Berlin Heidelberg, 2006, pp. 650--664.

\bibitem{GOLDSZTEJN2015105}
A.~Goldsztejn, C.~Jermann, V.~{Ruiz de Angulo}, and C.~Torras, ``Variable
  symmetry breaking in numerical constraint problems,'' \emph{Artificial
  Intelligence}, vol. 229, pp. 105--125, 2015.

\bibitem{DinechinPasca_DesigningCustomArithmetic_2011}
F.~de~Dinechin and B.~Pasca, ``{Designing Custom Arithmetic Data Paths with
  {FloPoCo}},'' \emph{{IEEE} Design {\&} Test of Computers}, vol.~28, no.~4,
  pp. 18--27, Jul. 2011.

\bibitem{KrachFrackeltonCarlettaVeillette_FPGAbasedimplementation_2003}
F.~Krach, B.~Frackelton, J.~Carletta, and R.~Veillette, ``{FPGA}-based
  implementation of digital control for a magnetic bearing,'' in
  \emph{Proceedings of the 2003 American Control Conference, 2003.}\hskip 1em
  plus 0.5em minus 0.4em\relax {IEEE}, 2003.

\bibitem{SarbisheiRadeckaZilic_AnalyticalOptimizationBit_2012}
O.~Sarbishei, K.~Radecka, and Z.~Zilic, ``{Analytical Optimization of
  Bit-Widths in Fixed-Point {LTI} Systems},'' \emph{{IEEE} Transactions on
  Computer-Aided Design of Integrated Circuits and Systems}, vol.~31, no.~3,
  pp. 343--355, Mar. 2012.

\bibitem{CarlettaVeilletteKrachFang_Determiningappropriateprecisions_2003}
J.~Carletta, R.~Veillette, F.~Krach, and Z.~Fang, ``{Determining appropriate
  precisions for signals in fixed-point IIR filters},'' in \emph{Proceedings
  2003. Design Automation Conference}, Jun. 2003, pp. 656--661.

\bibitem{DunningHuchetteLubin2017}
I.~Dunning, J.~Huchette, and M.~Lubin, ``Jump: A modeling language for
  mathematical optimization,'' \emph{SIAM Review}, vol.~59, no.~2, pp.
  295--320, 2017.

\bibitem{CPLEXUsersManual_2020}
\BIBentryALTinterwordspacing
CPLEX, ``{CPLEX User's Manual},'' 2020. [Online]. Available:
  \url{https://www.ibm.com/analytics/cplex-optimizer}
\BIBentrySTDinterwordspacing

\bibitem{VolkovaHilaireLauter_ReliableEvaluationWorst_2015}
A.~Volkova, T.~Hilaire, and C.~Lauter, ``{Reliable Evaluation of the Worst-Case
  Peak Gain Matrix in Multiple Precision},'' in \emph{2015 {IEEE} 22nd
  Symposium on Computer Arithmetic}.\hskip 1em plus 0.5em minus 0.4em\relax
  {IEEE}, Jun. 2015.

\end{thebibliography}


\begin{IEEEbiography}[{\includegraphics[width=1in,height=1.25in,clip,keepaspectratio]{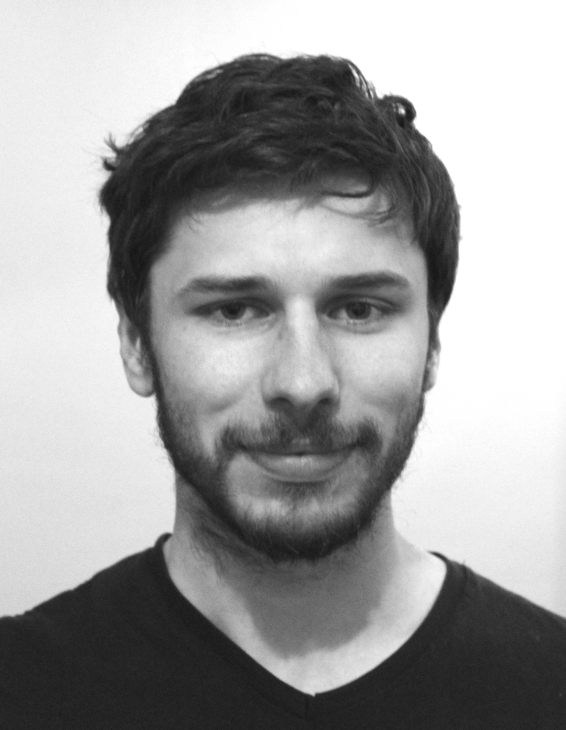}}]{R\'{e}mi Garcia} obtained his Bachelor's Degree in Mathematics and his Master's Degree in Computer Science from University of Nantes, France, in 2016 and 2020, respectively. He joined the LS2N, France, in 2020 as a PhD student in Computer Science. His research interests include optimization applied to signal processing and computer arithmetic.
\end{IEEEbiography}

\begin{IEEEbiography}[{\includegraphics[width=1in,height=1.25in,clip,keepaspectratio]{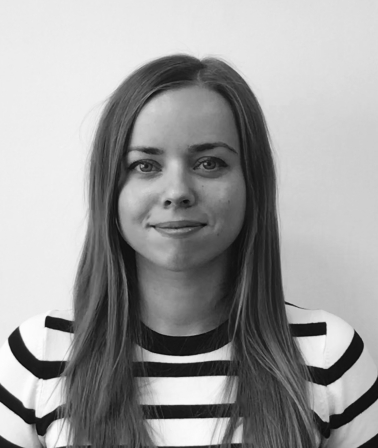}}]{Anastasia Volkova} received her Master’s Degree in Applied Mathematics from
Odessa National University, Ukraine, in 2014. She obtained a PhD in Computer Science from Sorbonne
University in Paris, France in 2017. She was a postdoctoral researcher at Inria, France and an AI
research resident at Intel Corporation. In 2019 she joined University of Nantes, France, as an Associate
professor. Her research interests include computer arithmetic, validated numerical computing and design
of optimized software/hardware for floating-point and fixed-point~{algorithms}.
\end{IEEEbiography}

\begin{IEEEbiography}[{\includegraphics[width=1in,height=1.25in,clip,keepaspectratio]{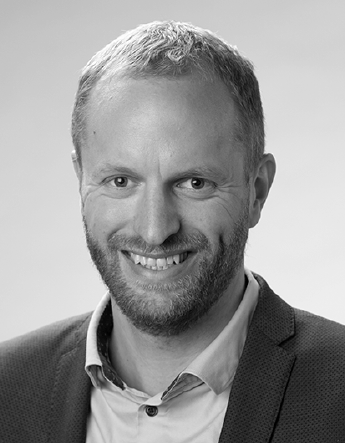}}]{Martin Kumm} received the Dipl.-Ing. degree in electrical engineering from the
University of Applied Sciences Fulda, Germany, and the Technical University of Darmstadt, Germany, in
2003 and 2007, respectively. From 2003 to 2009, he was with GSI Darmstadt, working on digital RF
control systems for particle accelerators. In 2015 he received his Ph.D. (Dr.-Ing.) degree from the
University of Kassel, Germany. He is currently a Professor for Embedded Systems at the Fulda University
of Applied Sciences, Germany. His research interests are arithmetic circuits and their optimization as well
as high-level synthesis, all in the context of reconfigurable systems.
\end{IEEEbiography}

\begin{IEEEbiography}[{\includegraphics[width=1in,height=1.25in,clip,keepaspectratio]{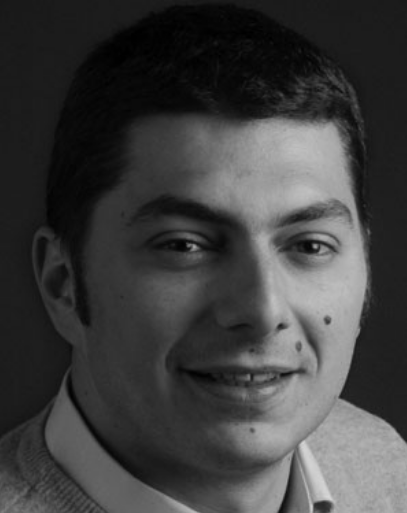}}]{Alexandre Goldsztejn} received the Engineer degree in computer science and mathematics from the Institut Sup\'{e}rieur d’Electronique et du Num\'{e}rique, Lille, France, in 2001, and the Ph.D. degree in computer science from the University of Nice Sophia Antipolis, Nice, France, in 2005. He has spent one year as a Postdoctoral Fellow with the University of Central Arkansas, Conway, AR, USA, and the University of California, Irvine, CA, USA. Since 2007, he has been a full-time CNRS Researcher with the Laboratoire des Sciences du Num\'{e}rique de Nantes, Nantes, France. His research interests include interval analysis and its applications to constraint satisfaction, nonlinear global optimization, robotics, and control.
\end{IEEEbiography}

\begin{IEEEbiography}[{\includegraphics[width=1in,height=1.25in,clip,keepaspectratio]{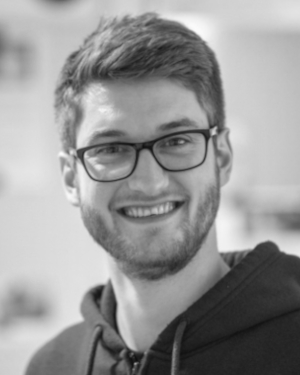}}]{Jonas K\"{u}hle} received his Bachelor's degree in Applied Computer Science from University of Applied Sciences Fulda in 2017
where he is about to complete his Master's degree in 2021 and will start as a PhD student in 2022.
\end{IEEEbiography}
\vfill

\end{document}